\documentclass[a4paper,fleqn,useAMS,usenatbib]{mnras}

\pdfoutput=1

\usepackage{natbib}
\usepackage{mathptmx}

\usepackage[T1]{fontenc}
\usepackage{ae,aecompl}

\usepackage[dvips]{graphicx}
\usepackage{amsmath}
\usepackage{amsfonts}
\usepackage{amssymb}
\usepackage{comment}
\usepackage{bm} 
\usepackage[dvipsnames]{xcolor}
\usepackage[]{hyperref}
\hypersetup{
	colorlinks=true,	
	urlcolor=MidnightBlue,		
	hypertexnames=true,
	plainpages=false,
	naturalnames=true,
	pdftitle={Astrometric Identification of Nearby Binary Stars I},    
	pdfauthor={Zephyr Penoyre},     
	linkcolor=WildStrawberry,          
	citecolor=Periwinkle,        
}

\usepackage{amsmath}
\usepackage{amsfonts}
\usepackage{amssymb}
\usepackage{wasysym}
\usepackage[percent]{overpic}
\usepackage{subfig}
\usepackage{csvsimple}
\usepackage{lscape}
\usepackage[shortlabels]{enumitem}

\makeatletter
\def\env@matrix{\hskip -\arraycolsep 
  \let\@ifnextchar\new@ifnextchar
  \array{*{\c@MaxMatrixCols}c}}
\makeatother

\newcommand{\cz}{\mathrm{c}}
\newcommand{\sz}{\mathrm{s}}

\title[Astrometric Identification of Binaries I]{Astrometric Identification of Nearby Binary Stars I: Predicted Astrometric Signals}
\author[Z. Penoyre et al.]{Zephyr Penoyre$^{1}$\thanks{E-mail:
\href{mailto:zpenoyre@ast.cam.ac.uk}{zpenoyre@ast.cam.ac.uk}}, Vasily Belokurov$^{1}$, N. Wyn Evans$^{1}$ \\
$^{1}$Institute of Astronomy, University of Cambridge, Madingley Road, Cambridge, CB3 0HA, United Kingdom}
\date{Accepted . Received ; in original form 9 2021 August 19}

\pubyear{2021}

\begin{document}
\label{firstpage}
\pagerange{\pageref{firstpage}--\pageref{lastpage}}
\maketitle

\begin{abstract}
We examine the capacity to identify binary systems from astrometric errors and deviations alone. Until the release of the fourth \textit{Gaia} data release we lack the full astrometric time series that the satellite records, but as we show can still infer the presence of binaries from the best fit models, and their error, already available. We generate a broad catalog of simulated binary systems within 100 pc, and examine synthetic observations matching the \textit{Gaia} survey's scanning law and astrometric data processing routine. We show how the Unit Weight Error (UWE) and Proper Motion Anomaly (PMA) vary as a function of period, and the properties of the binary. Both UWE and PMA peak for systems with a binary period close to the time baseline of the survey. Thus UWE can be expected to increase or remain roughly constant as we observe the same system over a longer baseline, and we suggest $UWE_{eDR3}>1.25$ and $\Delta UWE/UWE_{eDR3}>-0.25$ as criteria to select astrometric binaries. For stellar binaries we find detectable significant astrometric deviations for 80-90\% of our simulated systems in a period range from months to decades. We confirm that for systems with periods less than the survey's baseline the observed $UWE$ scales $\propto \ \varpi$ (parallax), $a$ (semi-major axis) and $\Delta =\frac{|q-l|}{(1+q)(1+l)}$ where $q$ and $l$ are the mass and light ratio respectively, with a modest dependence on viewing angle. For longer periods the signal is suppressed by a factor of roughly $\propto P^{-2}$ (period). PMA is largest in orbits with slightly longer periods but obeys the same approximate scaling relationships.
\end{abstract}

\begin{keywords}
astrometry,
parallaxes,
proper motions,
binaries: general
\end{keywords}

\section{Introduction}

Two sources of light, sufficiently distant from an observer with an aperture of a given size, merge into one apparent point source. If one is significantly brighter, its luminosity and colour dominate, and the presence and properties of the other are diluted or obscured. In many ways, the system looks indistinguishable from a single source. To detect the companion, we must look for telltale idiosyncrasies, inconsistent with the properties of a single star.

In \citet{Penoyre20} (P+20), we modelled how the motion of the centre of light of a binary system, around its centre of mass, deviates from motion consistent with a single source. If we assume every point source is a single star and fit a corresponding  model\footnote{Described by 5-parameters: the position on-sky at some reference epoch, $\alpha*$ \& $\delta$, the linear proper-motion, $\mu_\alpha*$ \& $\mu_\delta$, and the parallax $\varpi$.}, then binary systems will give particularly poor convergence, parameterised by the Unit Weight Error (UWE). In \citet{Belokurov20} (B+20), we examined populations of point sources in the second data release of the \textit{Gaia} survey (DR2, \citealt{Gaia16,Gaia18}), showing that we can use excess astrometric error to pick out higher proportions of binaries for particular types of source (Blue Stragglers, Blue Horizontal Branch stars and Hot Jupiter hosts) and as a function of position on the Hertzsprung-Russell (HR) diagram.

However, while we were able to make successful inferences about populations of stars, we were cautious to extend our analysis down to individual sources. Given that our telltale of binary behaviour is a particularly poorly converged fit, we might expect some unknown amount of contamination by other sources of noise injection, either astronomical or systematic.

The third astrometric data release of the \textit{Gaia} mission (eDR3, \citealt{Gaia16,Gaia21}) has extended the possible observational time bas of point sources from 22 months in DR2 to 34, providing similar data products with increased precision. Each observation is processed anew with each data release (sometimes being re-assigned to another source) and thus the inferred properties of stars can change and improve both due to more data and improved data-handling and systematics.

In this series of two papers we examine the behaviour of the astrometry of binaries in eDR3 data. In Paper I, we simulate systems and generate synthetic observations and fits as closely matched to the actual \textit{Gaia} astrometric pipeline as possible. This will inform us about which systems we expect to observe, how they are likely to present themselves and where our observations are most reliable. In paper II, we will examine the actual \textit{Gaia} eDR3 results, and the corresponding DR2 data. We will focus specifically on stars in the Gaia Catalog of Nearby Stars \citep{Smart20} and identify a large number of binary candidates from astrometry alone, as well as detailing methods to reduce and eliminate some contaminants.

Having two epochs of data allows us to examine the \textit{change} in the quality of astrometric fits, and to test if the behaviour of a point source remains consistent with excess motion caused by a binary. For example, if a source has a poor fit in DR2, but it is significantly improved in eDR3, we may conclude that the cause was one or two spurious data points, whilst if the fit is consistently bad, we can be increasingly confident that the cause is a second gravitating companion.

Thus, in this paper, we aim to set out and justify a minimal criteria to separate likely binary systems from both single stars and spurious astrometric solutions. We begin, in Section \ref{methodCompare}, with a general summary and comparison of available methods for the detection of binaries. In Section \ref{syntheticbinaries}, we generate a synthetic population of binaries and associated astrometric tracks. We fit these with an emulation of the Gaia astrometric pipeline to explore the behaviour of the inferred motion and goodness of fit as a function of the system properties. In Section \ref{changePeriod}, we dissect astrometric changes as a function of binary period, and in Section \ref{changeProperties} we break down the dependence on other binary properties. In Section \ref{changeObservable}, we specifically compare metrics which are observable in the current \textit{Gaia} data and finally in Section \ref{changeDetectability} we show what fraction of systems we would expect to have detectable excess $UWE$ and thus be identifiable in eDR3.

\section{A brief comparison of methods for binary identification}
\label{methodCompare}

Astrometric binaries probe a particular and relatively unexplored slice of the parameter space of multiple stellar systems. With the available Gaia DR2 and eDR3 data, we are able to identify systems with periods ranging from months to decades as we will show in Section \ref{changeDetectability}. This is longer than is observable in tight binaries or systems likely to show transits \citep{Stevens13}, whilst much shorter than most visually distinguishable wide binaries~\citep{ElBadry21}. Radial velocity measurements do cover this period range \citep{Wood21}, and form a natural comparison sample, but are limited to a small minority of stars (mostly due to the cost of individual measurements).

In P+20 we showed that astrometric identification is most sensitive to binaries with periods up to the length of time spanned by the dataset - in the case of Gaia, periods less than $\sim$ 10 years. If only a fraction of an orbit is observed, the extra motion is approximately linear and can be subsumed into proper motion. Thus, the signal is suppressed for systems with periods $\gtrsim$ 10 years. Shorter period orbits have smaller semi-major axes and thus the astrometric deviation is smaller too. This puts a soft lower limit, on the order of months, on identifiable periods.

Wide binary systems (see for example \citealt{ElBadry21}) are those where both components are separately resolvable and have measured positions and velocities consistent with the two objects being gravitationally bound. Observations are limited directly by their angular separation on-sky. Thus, larger separations and longer periods ($\gtrsim$ 10 years) are strongly preferred. 

Tight binary systems are those whose orbits are sufficiently close to significantly tidally distort the star and cause  measurable variation in the light curve (see for example \citealt{KS21,Prsa21}) - and occasionally are even aligned such that the stars eclipse each other. Most known systems are thus limited to very small orbits and short periods (<1 month, \citealt{Pawlak16,Soszynski16}). Transits are possible for wider binaries but the probability of the exact necessary alignment needed decreases precipitously for larger orbits. Compounding this is the need for multiple detections (and thus measurements over multiple orbital periods) to confirm detections.

Photometric detection, where the excess light from the companion is discernible in the combined spectra of the system (see for example \citealt{Milone08}), is independent of period and separation. However it is strongly limited to systems where both stars are similar in luminosity (or very distinct in temperature). The dimmer source must be detectable above the inherent uncertainty in the spectra of the brighter.

Finally, we can also detect spectroscopic binaries, where absorption lines in the spectra can be seen to shift over time due to the line of sight motion caused by the star's orbit (for example \citealt{Price-Whelan20}). This can be sensitive to a very wide range of periods. They are limited at long periods by the span of time over which observations have been taken. However it is still a relatively costly process, and multiple precise measurements exist only for a minority of stars. It also strongly favours stars with deep and easily identifiable spectral features, particularly evolved stars and the most massive end of the main sequence.

For both photometric and spectroscopic binaries the properties of the orbit can be directly inferred - giving an approximate period, eccentricity and some (often degenerate) insight into the mass ratio and inclination of the system.  Observing the same system with multiple methods is a powerful tool to break some of these degeneracies and is rapidly becoming viable for a widening selection of systems.

Binary properties can also be inferred by a fit to astrometric data. Eventually this will be possible with full Gaia epoch astrometry from the fourth data release onwards. Currently we must work from the astrometric noise and 5-parameter solution alone, from which only degenerate combinations of these parameters can be inferred.

The methods highlighted here have been chosen for their ubiquity, however many other detection methods exist. In particular there are many which require one or both of the companions to be of a particular stellar type or stage in its life cycle. We encourage interested readers to refer to \citet{Moe17} and references therein for a fuller description of the breadth of available methods.

It is only by observing a system with some or all of these methods that we can rule out companions with any certainty. A detection with one method does not exclude the detection of other components on very different orbits with other techniques \citep{Wood21}. All the above methods (including astrometry) will miss companions of sufficiently low mass ratio ($q$) and many also fail for low light ratio ($l$). We cannot exclude, and may even expect, some massive planets, brown dwarfs or compact objects residing in these systems which remain undetectable for the near future.

\subsection{Higher multiples}

Finally we should note that for many of the methods listed above, including astrometry, systems with higher multiplicities are also observable. Each added companion introduces a new relevant period - which will in general be quite distinct. 

Systems with more than two companions are only stable over significant timescales if they are arranged hierarchically, or exist in a resonance preserving the dynamical stability. For outer companions the inner system must be sufficiently compact to be dynamically indistinguishable to a single point mass. When all relevant periods are distinct, multiple methods can be used to identify both inner and outer companions. However, pathological cases can exist where the relevant periods are close (as could be the case, for example, with a four body system with two bound tight binary systems of similar periods). The signal from multiple pairs can interfere, making a clean inference of the presence and properties of the binaries much harder to discern.

Aliasing of relevant periods, due to the sparseness in time of \textit{Gaia} measurements, as well as astrophysical noise, e.g. from starspots, can significantly effect binaries but are a yet more confusing contaminant for higher multiple systems.

Throughout this paper we will talk only of binaries, but most of the logic and empirical results extends to systems where one or both of the point sources is itself a tight multi-body system. The same applies for systems where wider orbiting companions are also plausibly present but which contribute negligibly to the astrometric signal on the timescales of interest.

\section{Synthetic binaries}
\label{syntheticbinaries}

\begin{figure*}
\centering
\includegraphics[width=0.98\textwidth]{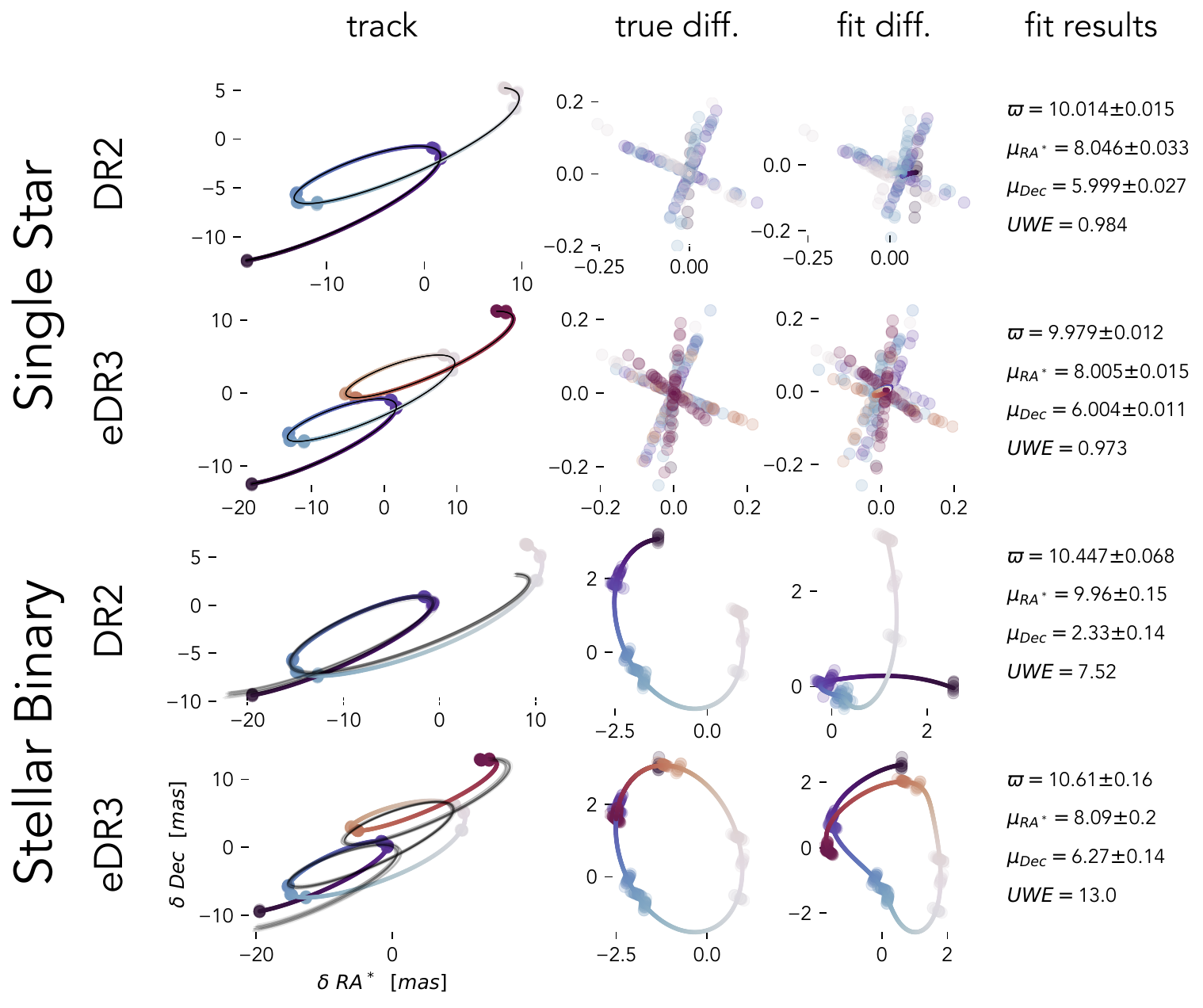}
\caption{An example of astrometric tracks and Gaia-like fits for DR2 and eDR3. We show the same system, as a single star (upper half) and with an unresolved stellar companion (lower half). The left hand column shows the astrometric track, coloured by time, from blue at the start of the Gaia survey to red at the end of the eDR3 observing time baseline. We show the whole track as a coloured line, and the simulated observations as large circles. We fit single-body 5-parameter astrometric tracks to the data using an emulation of Gaia's own astrometric pipeline, and a selection of these are shown in black. As we might expect, these only deviate from the actual track significantly in the case of a binary. In the middle columns we subtract a single body model to show the residuals - on the left hand side we use the true centre of mass motion, and on the right we use the best single body fit. On the far right we show the results from the single body fit - in particular the parallax ($\varpi$), proper motions ($\mu$) and unit weight error ($UWE$). Throughout we use $RA\cos(Dec)$ such that we are working in orthonormal coordinates. The mass of the primary is $1 M_\odot$, with $(RA,Dec)=(20,20)$ [degrees] and the scan angles and times are consistent with Gaia's sampling of those coordinates. The astrometric properties are $\varpi=10$ [mas], and $\bm{\mu}=(8,6)$ [mas yr$^{-1}$].  For the stellar binary $q=0.8$ and $l=q^{3.5}\approx0.46$. It has $P=2$ [yr], an initial phase of 0, $e=0.5$, and $(\theta_v,\phi_v,\omega_v)=(\frac{\pi}{4},\frac{\pi}{4},\frac{\pi}{4})$ [rad]. We use a flat 0.1 mas astrometric error throughout.}
\label{tracksexample}
\end{figure*}

\begin{figure}
\centering
\includegraphics[width=0.98\columnwidth]{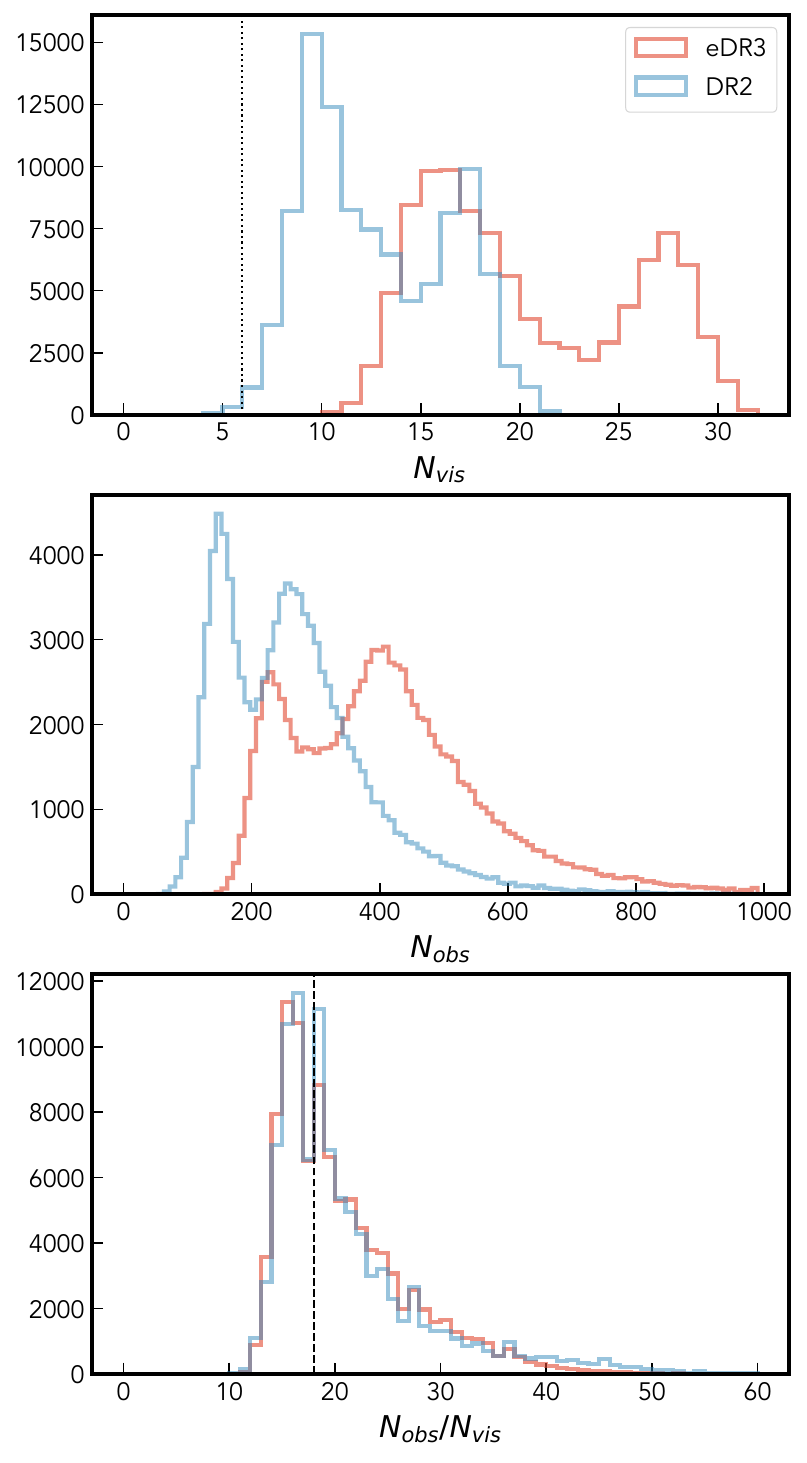}
\caption{(Top) Histogram of the number of visiblity periods (observations spaced out by more than 4 days) over the DR2 and eDR3 baselines for our 100,000 simulated systems. The bimodality shows the difference between the well sampled poles of the Gaia scanning law (oriented towards and away from the LMC) and the rest of the sky. Gaia only fits 5 parameter solutions to systems with 6 or more visibility periods (solid dotted line) and so the minority of systems which do not reach this criteria in DR2 are excluded from the following analysis. (Middle) Histogram of the total number of CCD observations of each system. (Bottom) Histogram of the average number of observations per visit for each system. We show a dashed line at 18, corresponding to two full sets of observations, one from each field of view, of 9 observations each.}
\label{simulated_nvis}
\end{figure}

In this paper we will present a synthetic population of binaries, with simulated observations which broadly mimic the capacity of the Gaia telescope. To this end we have developed a lightweight open source code \texttt{astromet.py}\footnote{\href{https://github.com/zpenoyre/astromet.py}{github.com/zpenoyre/astromet.py}} which simulates the position of the centre of light of a binary or single star system as a function of time. It also includes the functionality to fit the subsequent astrometric observations to a single star model through an emulation of Gaia's own single-body astrometric fitting pipeline \citep{Lindegren12}. Thus we can reproduce approximately the parameters and precision with which Gaia would observe such a system. This software is presented in more detail in the forthcoming paper Penoyre et al. 2022 (in prep.).

We randomly generated 50,000 systems, with parameters chosen from the random distributions in table \ref{tabRandom}. After generating the orbital parameters, we test each system with 5 different companions: a second main sequence star, black holes of 2, 5 and 10 $M_\odot$, and finally no companion at all. As the period is fixed, changing the secondary equates to changing only the semi-major axis of the orbit and the relative distance between the centre of mass and centre of light.

We adopt a convention of labelling the brighter of the binary the primary, thus for the black hole simulations the light ratio $l = \frac{L_B}{L_A} = 0$, whilst the mass ratio $q = \frac{M_B}{M_A} \gtrsim 1$ where $A$ and $B$ refer to the primary and secondary components of the binary. For the main sequence binary, we draw the mass of the secondary from the same distribution as the primary, then use a simple heuristic, $l=q^{3.5}$, to map the mass of a main-sequence star to its luminosity\footnote{This relationship between mass and light ratio, $l=q^{3.5}$ is a commonly used heuristic for main-sequence stars. More realistic relationships will vary as a function of mass of the primary, and all have some inherent scatter. See for example \citet{Kuiper38,Eker15,Lamers2017} for more detailed discussions.}. If the secondary mass is higher than the primary we swap the labels to ensure $l < 1$ always. The single star solutions always have $l=q=0$.

An example of the astrometric tracks and best fitting solutions is shown for a single star and a stellar binary in Figure \ref{tracksexample}. As well as the tracks and best fits we show the difference between single body models and the observations. For the correct center of mass motion this should always be an ellipse, but as the binary biases the astrometric solution the difference between the fit and the data is a more complex shape. We can also see here the effect of the scanning law, both in the irregular distribution of times in which the system is observed, and the scan angles. Whilst both binary fits significantly deviate from a single body motion, we see that the fit is much better over the shorter $DR2$ baseline. A lucky solution almost manages to pass through the small number of data points taken over that baseline.

Some of the parameter distributions, particularly $M_*, P$ \& $e$, are simplistic and approximate - partially due to the limited sample of known binaries in the range of periods of interest. The $M_*$ distribution comes from approximations of the present-day mass function locally \citep{Chabrier05}. The period distribution is chosen predominately to lie within periods in which we expect to see any astrometric effects (from a day to a century) evenly sampled in log space. In the future, we may be able to infer these distributions, as a function of period/stellar type/age or metallicity, by matching models of astrometric binaries to observed data. 

The distribution of proper motions come directly from Gaia eDR3 data, though it's worth noting that a Gaussian significantly underestimates the number of high velocity stars. The effect of this should be minimal as proper motion only introduces a linear component that separates easily from the parallax and binary orbit parameters. Only the angles and parallaxes have distributions which follow analytically from basic assumptions (isotropic orientations and constant local density of stars).

\begin{table}
\begin{tabular}{c c c}
 variable & description & distribution \\
 \hline
 $\varpi \ [mas]$ & Parallax & $10 \cdot \mathcal{U}[0,1]^{-\frac{1}{3}}$ \\
 $RA_0 \ [deg]$ & Right Ascension (t=0) & $360 \cdot \mathcal{U}[0,1]$ \\
 $Dec_0 \ [deg]$ & Declination (t=0) & $\frac{180}{\pi}\sin^{-1}(\mathcal{U}[-1,1])$ \\
 $\mu_{RA^*} \ [mas/yr]$ & RA cos Dec proper motion & $\varpi \cdot \mathcal{N}(0,6.67)$ \\
 $\mu_{Dec} \ [mas/yr]$ & Dec proper motion & $\varpi \cdot \mathcal{N}(0,6.67)$ \\
 \hline
 $P \ [yr]$ & Binary period & $10^{\mathcal{U}[-2.56,2]}$ \\
 $l$ & Binary luminosity ratio & \textit{see text} \\
 $q$ & Binary mass ratio & \textit{see text} \\
 $M_1 \ [M_\odot]$ & Mass of brightest component & $10^{\mathcal{N}(-0.65,0.6)}$ \\
 $t_0$ & Time of binary periapse & $P \cdot \mathcal{U}[0,1]$ \\
 $e$ & Eccentricity & $\mathcal{U}[0,1]$ \\
 $\theta_v \ [rad]$ & Polar viewing angle & $\cos^{-1}(\mathcal{U}[-1,1])$ \\
 $\phi_v \ [rad]$ & Azimuthal viewing angle & $2 \pi \cdot \mathcal{U}[0,1]$ \\
 $\omega_v \ [rad]$ & Planar projection angle & $2 \pi \cdot \mathcal{U}[0,1]$ \\
\end{tabular}
\caption{Parameters and distributions used to generate the mock observations. The first five define the single body astrometric motion and the others define the binary motion. Here $\mathcal{U}[a,b]$ represents a uniformly drawn random number between $a$ and $b$, $\mathcal{N}(\mu,\sigma)$ represents a number drawn from a normal distribution with mean $\mu$ and width $\sigma$. The semi-major axis of the orbit is another necessary element but can be calculated from $P$, $M_1$ and $q$.}
\label{tabRandom}
\end{table}


We use an approximation of the true eDR3 Gaia scanning law \citep{Green2018,Boubert20} from the \texttt{scanninglaw}\footnote{\href{https://github.com/gaiaverse/scanninglaw}{github.com/gaiaverse/scanninglaw}} package, which returns, as a function of on-sky position, the times and associated scanning angles at which Gaia would observe a source. Like in the published Gaia DR2 and eDR3 astrometric analyses, we only use the one-dimensional along-scan data. The error on across scan measurements is significantly (at least 5 times) larger and thus would anyway contribute negligibly to the fit.

Though we will use eDR3 to label our data throughout this paper the astrometric data in the full third data release (DR3) should be identical to that in eDR3. The choice of which label to use here is an arbitrary convention.

The number of visibility periods (observations spaced by at least 4 days) is shown in Figure \ref{simulated_nvis}. This is a good proxy for the number of independent observations, and we can see the clear improvement in coverage from DR2 to eDR3 (22 months to 34 months). In general there are 9 (CCD) observations within a single transit of the telescope\footnote{To be exact, one of the 7 rows of CCDs Gaia uses for astrometric measurements has only 8 detectors, thus the average number of observations per transit is actually $8\frac{7}{8}$, though we don't include that detail here.}. Similarly for many sources both of Gaia's fields of view will pass over the source in quick succession, thus there may be 18 or more measured observations corresponding to a single visibility period. This means realistic observations, such as we simulate here, have very tightly bunched groups of astrometric measurements, unlike the random distribution used previously in P+20. We can see this examining the middle and bottom panels of the figure - the total number of observations is much higher and on average 18 or more measurements are taken for each visibility period, as predicted in \citet{Eyer17}.

One major departure from realistic observations is the use of a flat (along-scan) astrometric error, of 0.1 mas. In reality, the uncertainty varies greatly as a function of apparent magnitude, but this value was chosen to be within a factor of two of the expected value for a wide range of apparent magnitudes ($6\lesssim m_G \lesssim14$, from \citealt{Lindegren21} figure A.1). This makes our simulated results easier to interpret, but perhaps unrealistic for particularly bright or dim sources. Note that this is the error per observation and thus by the above logic the error associated with a single visibility period may be expected to be an order of magnitude smaller ($\sim\sqrt{\frac{1}{18}}$).

We are interested in comparing observations from two windows of time. It is important to note the distinction between observations such as these, where the time baselines over which the system is observed overlap, as opposed to observations of a system over two completely distinct time periods. The current and future Gaia data releases will always contain data derived from observations that were also used in the earlier release, thus each Gaia data release solution is not an independent measure of the same system at different epochs.

A similar set of simulations was presented in P+20, and a broad discussion of the effect of orbital properties on observed $UWE$ can be found there. Here, we focus in particular on the effects found when comparing two observing time baselines.

\section{Astrometric change as a function of period}
\label{changePeriod}

We start by examining the change in  astrometric error ($UWE$), proper motion ($\bm{\mu}$) and parallax ($\varpi$) and how they depend on binary \textit{period}. Whilst it is not a directly observable quantity, it is the period which is most important in separating different populations of binaries.

\subsection{Change in $UWE$}
\label{changeInUwe}

\begin{figure*}
\centering
\includegraphics[width=0.99\textwidth]{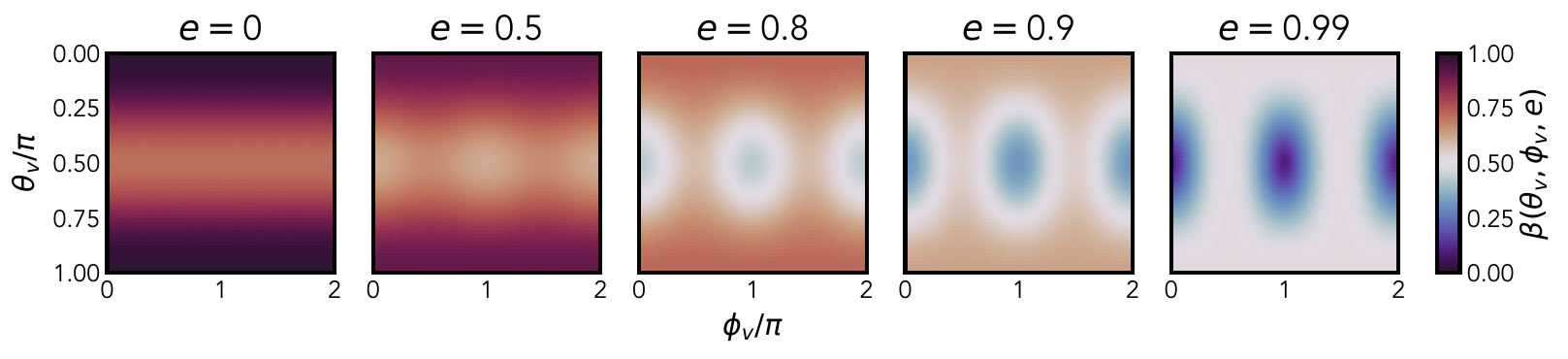}
\caption{The variation in the projection term, $\beta(\theta_v,\phi_v,e)$, of the average astrometric deviation caused by a binary (equations \ref{uwesimple} and \ref{betaTerm}). $\theta_v$ is the polar viewing angle (often called inclination) and is $\frac{\pi}{2}$ for a system viewed edge on and 0 or $\pi$ for a face-on system. $\phi_v$ is the azimuthal viewing angle, defined relative to periapse - thus the astrometric contribution of very eccentric systems goes to zero for $\phi_v=0$ or $\pi$, as all of the motion is directly along the line of sight.}
\label{projectionTerm}
\end{figure*}

\begin{figure}
\centering
\includegraphics[width=0.49\textwidth]{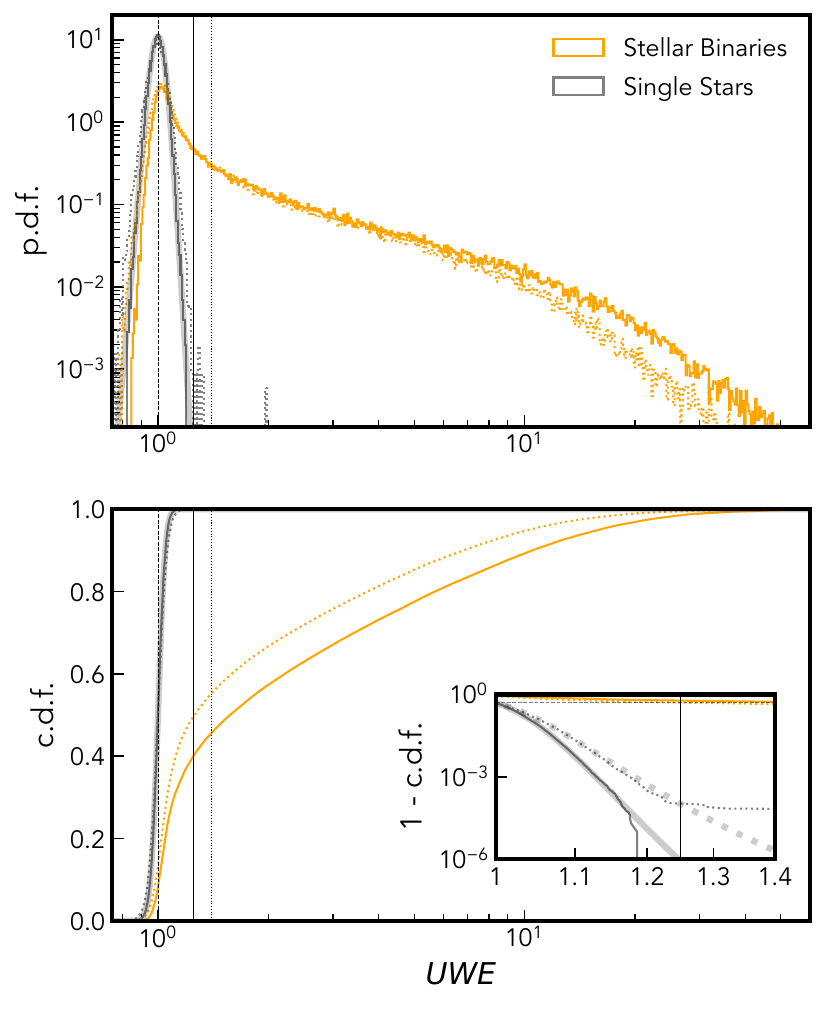}
\caption{The probability distribution function (p.d.f. - upper), and cumulative distribution function (c.d.f. - lower) of $UWE$ for single stars (grey) and stellar binaries (orange). Solid lines correspond to observations over the eDR3 baseline, and dotted show the DR2 values. The inset in the lower panel shows the survival function, the fraction of sources with $UWE$ greater than a given value. A dashed vertical line shows an $UWE$ of 1, whilst the solid and dotted lines show values of 1.25 and 1.4 respectively. We fit a student-t distribution to the single stars in eDR3 (see text for details) and this profile is shown as a thick grey line. The corresponding DR2 fit is also shown in the inset, as a dotted thick grey line.}
\label{uweDist}
\end{figure}

\begin{figure}
\centering
\includegraphics[width=0.49\textwidth]{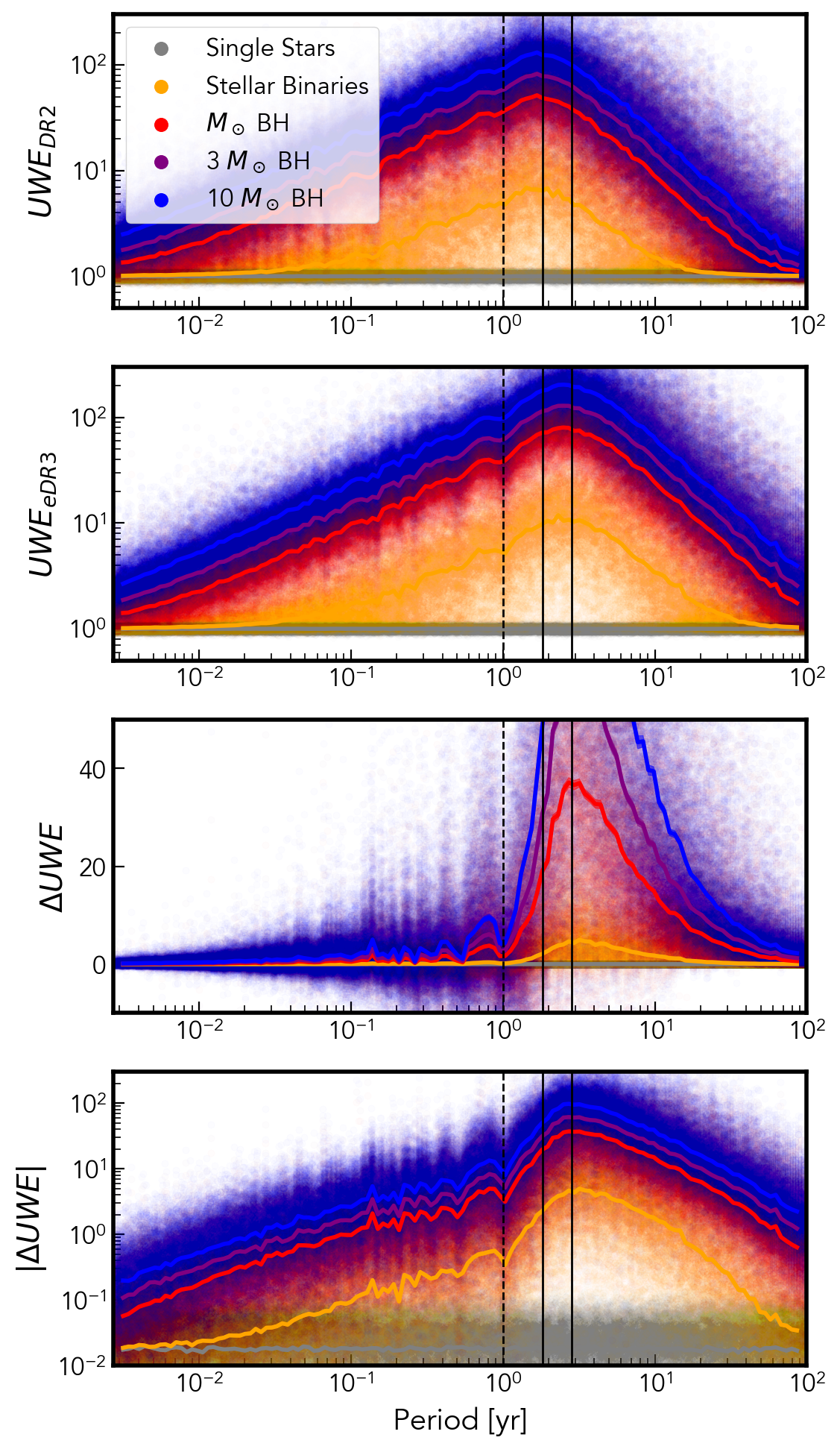}
\caption{UWE and change in UWE as a function of period for each population of simulated binaries. Each individual simulated system is shown as a translucent point, and the median for each population, as a function of period, is shown as a solid line. The dashed vertical line corresponds to a 1 year period, and the solid verticals show the DR2 and eDR3 observing baselines. The top two panels show the measured UWE over the DR2 and eDR3 baselines respectively, whilst the latter panels show the change in measured UWE and the absolute value thereof.}
\label{deltaUwe}
\end{figure}

\begin{figure}
\centering
\includegraphics[width=0.49\textwidth]{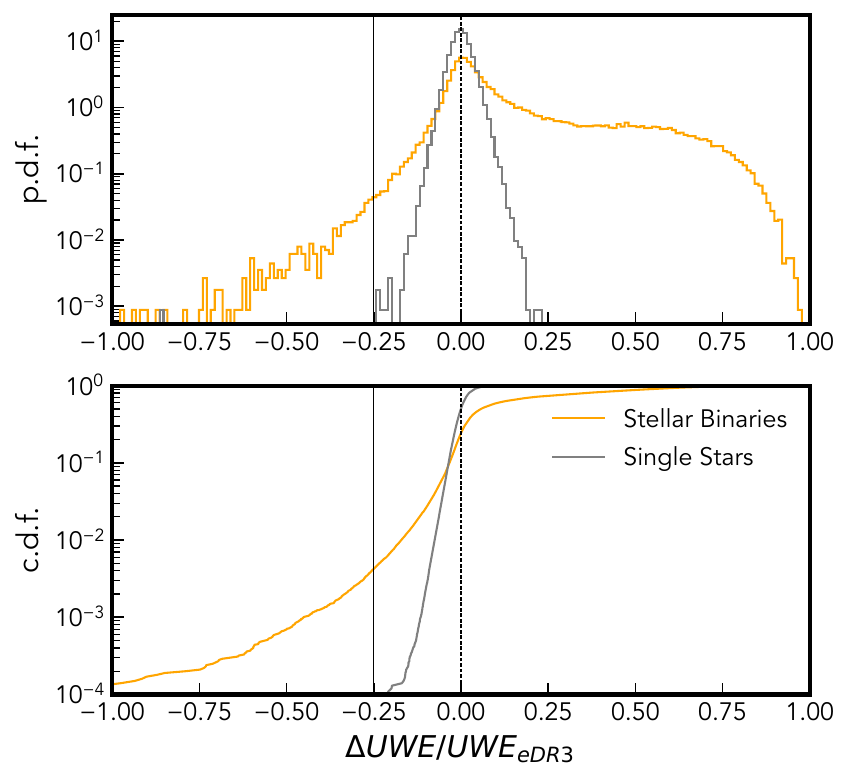}
\caption{The p.d.f. (upper) and c.d.f. (lower) of the change in $UWE$ between eDR3 and DR2, scaled by $UWE_{eDR3}$. Similarly to figure \ref{uweDist} we see that single stars have a relatively symmetric and tight distribution around the expected value of 0 (shown with a dashed vertical line). The distribution of binaries is much broader, still peaking at 0 but with a heavy tail of positive $\Delta UWE$ (with a maximum value of 1, for which $\frac{UWE_{DR2}}{UWE_{eDR3}}\rightarrow 0$) and a minority of sources with significant negative $\Delta UWE$. We suggest a critical $\frac{\Delta UWE}{UWE_{eDR3}}$ of -0.25 to some sources dominated by spurious astrometry whilst removing only a small fraction ($<1\%$) of true binaries.}
\label{relativeDeltaHist}
\end{figure}

The unit weight error ($UWE$)\footnote{The equivalent measure published in Gaia is the re-normalised unit weight error (RUWE), which has been rescaled to account for the unknown astrometric error, see \citet{Lindegren18}. In the second paper of this series we will later introduce the Local Unit Weight Error (LUWE) which renormalises UWEs over just the local region. In most respects it is safe to take them as interchangeable, though we will attempt to use $UWE$ primarily for theoretical predictions and $RUWE/LUWE$ for observational data.} is effectively the square root of the reduced chi squared value of the observed data compared to the best fitting single star astrometric model.

Imagine we have a series of $N$ observations, at times $t_i$, of a source with true position on-sky of $\alpha^*_i=RA_i \cos Dec_i - RA_0 \cos Dec_0$ and $\delta_i=Dec_i-Dec_0$ where $RA_0$ and $Dec_0$ are some reference coordinates (allowing us to make an approximately linear local expansion). Gaia has much higher astrometric precision along the direction it scans than across it, and thus for most cases it is sufficient to think of observations as being one-dimensional along some scan angle $\phi_i$ (relative to our co-ordinate system). Thus the recorded position is
\begin{equation}
x_i = \alpha^*_i\sin\phi_i + \delta_i\cos\phi_i + \mathcal{N}(0,\sigma_{ast})
\end{equation}
where we've assumed that measurement errors are normally distributed with standard deviation $\sigma_{ast}$.

We can compare this to the predictions of an astrometric model, for example the 5-parameter single-body solution, which predicts positions $\bar{\alpha}^*(t_i)$ and $\bar{\delta}(t_i)$. Using $\bar{x}(t_i)=\bar{\alpha}^*(t_i)\sin\phi_i + \bar{\delta}(t_i)\cos\phi_i$ we can calculate the chi-squared value
\begin{equation}
\label{chi2}
\chi^2=\sum_{i=0}^{N_{obs}} \frac{(x_i - \bar{x}(t_i))^2}{\sigma_{ast}^2}.
\end{equation}

If our model is good we expect $\chi^2$ to be close in value to the number of observations, minus the degrees of freedom. Thus the value
\begin{equation}
\label{UWEchi}
UWE=\sqrt{\frac{\chi^2}{N - 5}}
\end{equation}
should be close to 1 for a system well described by a 5 parameter single-body astrometric model. Values significantly less than one suggest an overestimated astrometric error, whilst large values can be caused by one of three factors:
\begin{itemize}
\item Multiple systems - The track of the centre of light of a binary (or higher multiplicity) system can deviate substantially from the simple single body astrometric solution. A second point mass introduces a second component of Keplerian elliptical motion (the first being the parallax ellipse) around which the centre of light moves over one binary period. This can cause discrepancies of up to $\sim$ 180 milli-arcseconds (mas), above which we enter the regime where Gaia may start to be able to resolve them seperately (though in practice sources up to a few arcseconds apart may still be imperfectly resolved). Additionally there can be a variety of temporary non-Keplerian disturbances, like astrometric microlensing events. 
\item Source misidentification - Each Gaia source is moving against a field of other points of light, each on its own track. If the position of one source is recorded as belonging to another, even in just a few of the many observations Gaia makes of each source, this can introduce significant excess noise, inconsistent with either a single star or a multiple system. This is most likely to happen when one of the two fields of view is pointed toward a crowded region, such as low galactic latitudes, or nearby clusters. It can also be more of a concern for fast-moving sources which sweep across a larger area of sky. We can also include close and blended sources in this category, which overlap on-sky for some or all of the Gaia observing window. Observations of these overlapping sources at different times may record the position as being that of either one of the sources or of the photocentre of both. Astrometry starts to deteriorate for two sources as far
from each other as 2 arcseconds (see B+20 for discussion).
\item Unmodelled noise - This is a \lq catch all' for all the other reasons a source may have an erroneous position recorded. This could be due to random changes in the source, the telescope, the data pipeline or one of a host of possible fringe causes. With a dataset so large some low degree of occasional random error injection is inevitable. This also includes the possibility of significant underestimation of the astrometric error, inflating the $\chi^2$ (equation \ref{chi2}).
\end{itemize}

It is worth noting that only in the case of a binary (or higher multiple) would we expect to see consistent deviations across time. Both misidentification and most forms of random error may cause occasional large deviations. Excess motion caused by an unresolved companion will be present across all observations, and the discrepancy becomes more significant the longer we observe. Outlying data points are down-weighted automatically by the Astrometric Global Iterative Solution (AGIS, \citealt{Lindegren12}) which will minimise these effects, but cannot entirely exclude them. 

The relative motion of the centre of light simply adds an extra elliptical component to the astrometric track. Its amplitude (compared to the random astrometric noise) depends on the size of the orbit, the distance of the system, and whether it is visible over astrometric noise. It has an orientation set by the viewing angle of the binary, and a syncopation of the motion set by the eccentricity.

If the astrometric offset from the centre of mass at time $t_i$ is $\bm{\epsilon_i}$ then the residual of a 1D observation at scan angle $\phi_i$ and the true centre of mass motion should follow
\begin{equation}
R_i = \mathcal{N}[0,\sigma_{ast}] + \begin{pmatrix} \sin \phi_i \\ \cos \phi_i\end{pmatrix} \cdot \bm{\epsilon_i}
\end{equation}
where $\mathcal{N}[0,\sigma_{ast}]$ is a normally distributed astrometric error. Thus
\begin{equation}
\label{resSq}
\begin{aligned}
\chi^2 = \frac{1}{\sigma_{ast}^2}\sum_{i}R_i^2 =&\frac{1}{\sigma_{ast}^2}\sum_{i} \big(\mathcal{N}^2 + 2 \mathcal{N} (\sin \phi_i \epsilon_{i,\alpha^*} + \cos \phi_i \epsilon_{i,\delta}) \\
&+ \sin^2 \phi_i \epsilon_{i,\alpha^*}^2 + \cos^2 \phi_i \epsilon_{i,\delta}^2 + 2 \sin \phi_i \cos \phi_i \epsilon_{i,\alpha^*} \epsilon_{i,\delta}\big).
\end{aligned}
\end{equation}
If the number of observations, $N$, is large then we expect $\sum \mathcal{N} = 0$ and $\sum \mathcal{N}^2 = N \sigma_{ast}^2$. If the scanning angles are approximately random $\sum \sin \phi_i = \sum \cos \phi_i = \sum \sin \phi_i \cos \phi_i = 0$ and $\sum \sin^2 \phi_i = \sum \cos^2 \phi_i =\frac{1}{2}$. Finally we can define an average angular separation between the centre of mass and centre of light
\begin{equation}
\delta \theta^2 = \frac{\sum_i |\bm{\epsilon_i}|^2}{N}.
\end{equation}
Thus under these assumption we can reduce equation \ref{resSq} to 
\begin{equation}
\chi^2 = N\left(1^2+\frac{\delta\theta^2}{2\sigma_{ast}^2}\right)
\end{equation}
and thus equation \ref{UWEchi} can be expressed as
\begin{equation}
\label{dtheta}
UWE=\sqrt{\frac{N}{N-5}}\sqrt{1+\frac{1}{2}\left(\frac{\delta\theta}{\sigma_{ast}}\right)^2}.
\end{equation}

$\delta \theta$ can also be translated into an average physical separation between the centre of light and centre of mass
\begin{equation}
\label{dsemi}
\delta a = \frac{\delta \theta}{\varpi}.
\end{equation}

In P+20 we showed that, for a binary with semi-major axis $a$ (in AU) and mass \& light ratios $q$ \& $l$, for which we observe the position approximately uniformly over one full period we expect the average angular deviation over one orbit should scale as 
\begin{equation}
\label{uwesimple}
\delta \theta =\ a \ \varpi \ \Delta(q,l) \ \beta(\theta_v,\phi_v,e)
\end{equation}
where
\begin{equation}
\label{deltaql}
\Delta(q,l) = \frac{q-l}{(1+q)(1+l)}
\end{equation}
is the relative distance of the centre of light of the system from it's centre of mass and
\begin{equation}
\label{betaTerm}
\beta(\theta_v,\phi_v,e)=\sqrt{1-\frac{\sin^2\theta_v}{2} - e^2 \frac{3+\sin^2\theta_v(\cos^2\phi_v-2)}{4}}
\end{equation}
describes the dependence on the viewing angle of the system. A more exact expression for the UWE for arbitrary known period will be presented in Penoyre et al. 2022 (in prep.)). 

In figure \ref{projectionTerm} we show the variation in $\beta$ with viewing angle and eccentricity. The value is always between 0 and 1. Only the most eccentric orbits, aligned such that all the motion is along the line of sight, have $\beta$ close to zero. For orbits with small to moderate eccentricities ($e\lesssim0.5$) the value is mostly set by the $\frac{\sin\theta_v^2}{2}$ term, with $\beta$ close to one for face-on systems and reduced by roughly $\sqrt{2}$ when viewed edge-on. In general edge-on systems will have smaller signals, though only in the most eccentric cases can $\beta$ dip below 0.5. The minimum possible value, when $\sin\theta_v$ and $\cos\phi_v$ are equal to 1 is $\sqrt{\frac{1-e^2}{2}}$.

Equation \ref{uwesimple} is derived assuming many uniformly spaced observations over one binary period. Let's define the number of orbits observed
\begin{equation}
N_{orb}=\frac{B}{P}
\end{equation}
where $B$ is the baseline of the survey (or more accurately the time between first and last observations). Our relationship holds well for $N_{orb}>1$, performing worst in this case when $N_{orb}=n+\frac{1}{2}$ for some integer $n$. As $N_{orb}\rightarrow \infty$ the result tends to the given expression.

However many systems will have less than one period observed ($N_{orb}<1$) and in these cases predictions from equation \ref{uwesimple} diverge from observed values. This is because a small arc of an orbit is approximately linear, and the effect of the binary is to change the proper motion (which can be parameterised in a 5-parameter single body model) and the $UWE$ is small. For a much more detailed exploration of this we refer to P+20 and the upcoming Penoyre et al. 2022 (in prep.). We suggest here that for systems with $N_{orb}<1$ a more appropriate rough estimate of $UWE$ is
\begin{equation}
\label{UWEadj}
UWE'=\sqrt{\frac{N}{N-5}}\sqrt{1+\frac{1}{2}\left(\frac{N_{orb}^2\delta\theta}{\sigma_{ast}}\right)^2}
\end{equation}
where the factor of $N_{orb}^2$ is suggested due to a reasonable fit with the observed behaviour, rather than any deeper physical argument. 

Whilst we've shown that for $N_{orb}\gtrsim1$ eccentricity tends to decrease the projection term $\beta$ (equation \ref{betaTerm}) and thus the overall binary contribution, the behaviour is more complex for longer period orbits. A high eccentricity means that much of the motion is confined to a short fraction of the orbital period. Thus these orbits, if viewed at the right phase, can have tracks incompatible with linear motion (and thus a large $UWE$) even when the period is long. It is still true that more eccentric systems can be more suppressed by projection effects, and thus at long binary periods both the largest and smallest $UWE$ measurements come from very eccentric orbits.

We can compare the average angular deviation over one orbit, $\delta \theta$, to the size of the parallax ellipse $\varpi$, whose orientation is set by the position on-sky and with uniform motion (ignoring the slight eccentricity of Earth's orbit, and thus \textit{Gaia's} orbit around L2). Thus, the ratio of the two components scales proportional to $a \Delta$ (with $a$ measured in AU). 

$\Delta$ can take any value between -1 and 1. As shown in B+20 using the example of two Main Sequence stars in a binary, it can be expected to vary from very small magnitudes (if one star dwarfs the other, or if they are very similar in mass and luminosity) to a maximum of around 0.2-0.3. For other types of binary larger values are possible, but this gives a general idea of the likely magnitude range.

The very constrained nature of the parallax ellipse means that binary motion can be detected even when $a \Delta$ is small, as long as $\frac{a \varpi \Delta}{\sigma_{ast}} \gtrsim 1$, i.e. the binary motion dominates over the error.

\subsubsection{Distribution of UWE}
Figure \ref{uweDist} shows the distribution of UWE values for simulated single stars and stellar binaries over the DR2 and eDR3 time baselines. The single stars clearly peak at 1, as we expect and have a relatively narrow distribution.

We fit Student's t-distributions, using \texttt{scipy.stats.t}, to the DR2 and eDR3 single stars, of the form
\begin{equation}
p.d.f.(x,\nu)=\frac{\Gamma\left(\frac{\nu+1}{2}\right)}{\sqrt{\pi\nu}\Gamma\left(\frac{\nu}{2}\right)}\left(1+\frac{x^2}{\nu}\right)^{-\frac{\nu+2}{2}}
\end{equation}
where $\nu$ is the degrees of freedom and
\begin{equation}
x=\frac{UWE - \mu}{\sigma}
\end{equation}
is a shifting and rescaling of the $UWE$. For the DR2 period we find $\mu=0.9988, \sigma=0.0438$ and $\nu=9.79$. For eDR3 we find $\mu=0.9993, \sigma=0.0354$ and $\nu=17.40$.

We can see, both from the figure and the fit, that the scatter of $UWE$ is significantly reduced in the longer baseline data. The mean of both distributions, $\mu$, are consistent with unity. The number of degrees of freedom is approximately the average number of visibility periods (a good proxy for the number of independent observations) as can be seen with reference to figure \ref{simulated_nvis}.

The significantly smaller spread in eDR3 UWE motivated us to use a new critical value over which astrometric deviations are considered inconsistent with a single star. Previously a value of 1.4 has been adopted, as used in P+20 and B+20. We show in the inset panel that this is approximately where only one single star in a million is removed from DR2. Applying a similar criteria to our eDR3 distribution we suggest $UWE_{eDR3}<1.25$ as a comparable criteria for stars astrometrically consistent with a single body solution.

Turning now to the binaries we see that the distribution also peaks near, although significantly above, 1. This means no criteria on $UWE$ will completely seperate single stars from multiples, as shown in P+20. However examining the cumulative distribution functions we can see that whilst a little over half of all binaries simulated have $UWE_{DR2}<1.4$, only 40\% have $UWE_{eDR3}<1.25$. This fraction will never tend to zero, small or very long binary orbits, or pathologically aligned systems, will always be able to masquerade as single stars. However it is clear longer baseline observations will continue to improve the completeness of astrometric binaries.

In Appendix \ref{sec:uweDist_properties} we explore this distribution further, split by source parallax and binary period.

\subsubsection{Change in UWE from DR2 to eDR3}

Figure \ref{deltaUwe} shows the measured UWE over the DR2 and eDR3 time baselines for our simulated systems. In the first two panels, we show that the UWE peaks at the observational time baseline of the survey, corresponding to the widest binaries for which we measure at least one complete orbit.

As discussed in detail in P+20, the best fitting solution for a longer period orbit is that in which much of the motion due to the binary is taken as a shift in the linear proper motion. Hence we see a decline in UWE for periods above the time baseline of the observations ($\leq$22 months for DR2 and $\leq$34 months for eDR3).

We also see a slight but significant decrease in UWE for periods which are harmonic with a year - the most noticeable at 1 year itself, then at $\frac{1}{2}$ and $\frac{1}{3}$ of a year. At these periods some of the binary motion can be fitted by a shift in parallax, leading to a better fit (lower $UWE$) but an erroneous distance measurement. We will discuss this more in section \ref{changePllx}.

For the black hole hosting systems increasing the mass changes only $\Delta$ and the semi-major axis, the former of which varies as $\frac{q}{1+q}=(1+\frac{1}{M_{BH}})^{-1}$. The stellar binaries are generally of lower mass than the black holes, and have non zero $l$ which will suppress slightly the UWE (especially for $q$ close to unity). As we would expect, the single stars show no significant UWE.

The final two panels of figure \ref{deltaUwe} show the change in UWE between the DR2 and eDR3 values
\begin{equation}
\Delta UWE= UWE_{eDR3}- UWE_{DR2}.
\label{deltaUWEEq}
\end{equation}
As expected from the above argument, this peaks at the eDR3 observing basline, and is large for systems with periods longer than $\sim 1$ year.

A feature that we will use when examining the sample from the Gaia survey itself is that for the majority of true binary systems UWE increases, and even those systems for which it decreases (with periods immediately below the DR2 observing baseline) the change is slight. For systems above the DR2 baseline, the UWE increases markedly. This would not be true of random errors (like source misidentification and unmodelled noise) where any error will increase the $UWE$ but the significance of that error will decrease over time. 

We show this in figure \ref{relativeDeltaHist}, where both single and binary stars have a spread of $\Delta UWE$ that contains few significant negative values - and thus cutting out sources with $\frac{\Delta UWE}{UWE_{eDR3}}$ retains almost all single stars and over 99\% of stellar binaries.

What we cannot show easily with out simulations is the types of system this cut is effective in removing - those subject to infrequent injections of noise. This could come, for example, from passing near another star for some number of observations - leading to some small fraction of the astrometric track being inconsistent with the rest. In their simplest form we would expect these shifts to occur at a constant rate, thus the likelihood of a single such event occurring in any observing window is just proportional to the time baseline of the survey $B$. As $B_{DR2}>B_{eDR3}-B_{DR2}$ we would probably expect more of such events to occur during the $DR2$ window, and afterwards for the $UWE$ to decrease as we accrue more observations and the blip becomes less significant. In other words, under these simple assumptions we would expect more of these spurious astrometric errors to have a negative $\Delta UWE$, and an $UWE_{eDR3}$ approaching 1, and thus the cut on $\Delta UWE/UWE_{eDR3}$ should be more effective at removing these than true binary systems.

Furthermore, a modest change in UWE, compared to the total UWE, suggests a system has a relatively short period - thus we can use the ratio of $\frac{\Delta UWE}{UWE}$ as a very basic proxy for the period. However we do not suggest that this gives a simple one-to-one mapping as the relative change will depend on other properties of the system. We discuss this more in section \ref{relDeltaProp}.

Longer observing baselines (as will be provided by future Gaia data releases) will further help constrain the true $UWE$ values of these systems. Similarly they will allow us to differentiate between binaries of differing periods.

\subsection{Proper motion anomaly}

\begin{figure}
\centering
\includegraphics[width=0.49\textwidth]{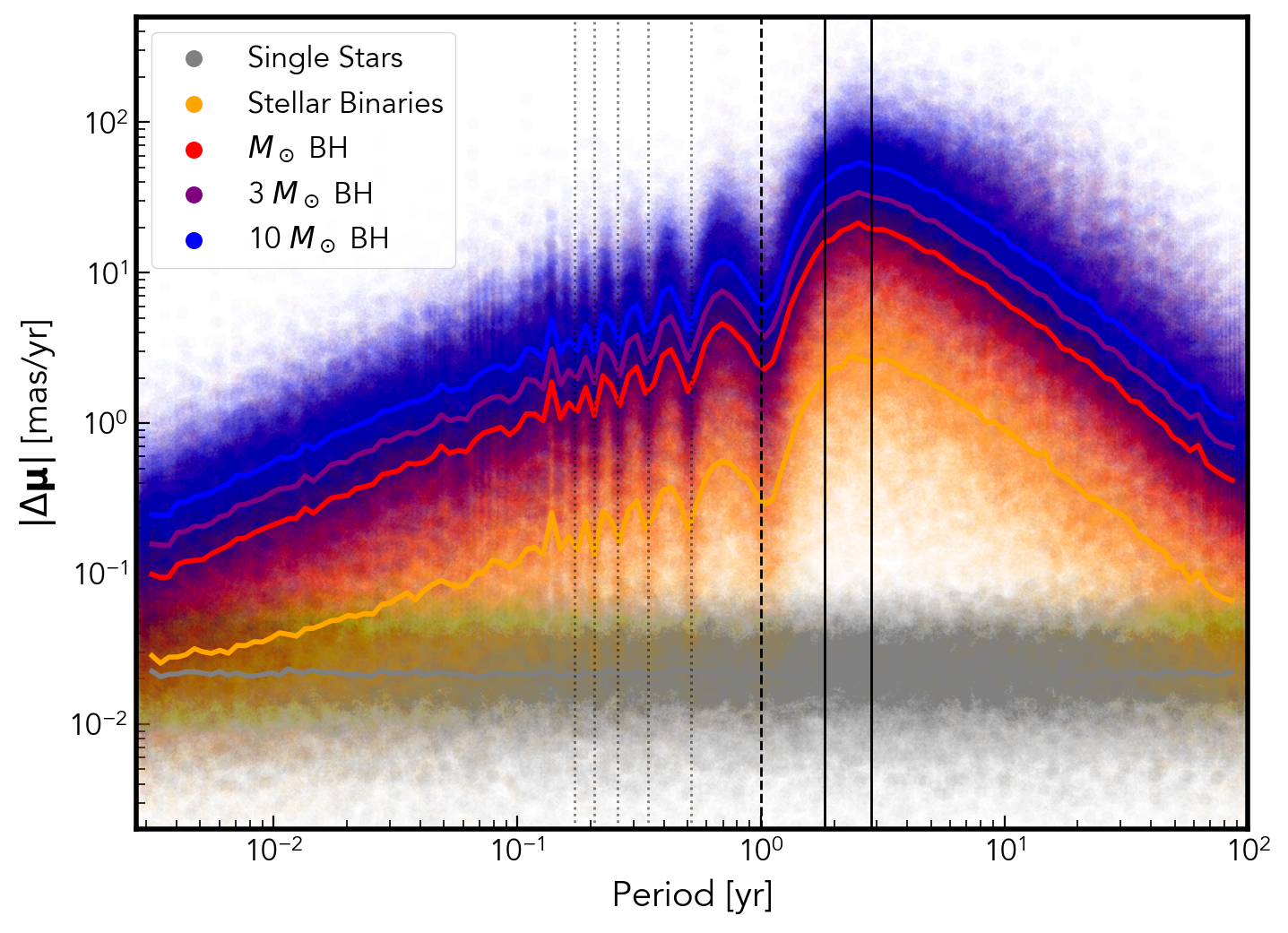}
\caption{Similar to figure \ref{deltaUwe} now showing the absolute value of the proper motion anomaly. Also shown as dotted lines are harmonics of \textit{Gaia}'s precession period of $P_{prec}=63$ days (moving from low period to high we show $P=n P_{prec}$ where $n=1,\frac{6}{5},\frac{3}{2},2,3$), as binaries of these periods complete a close to integer number of orbits over both the $DR2$ and $eDR3$ baselines and thus the proper motion anomaly is surpressed.}
\label{deltaMu}
\end{figure}


\begin{figure}
\centering
\includegraphics[width=0.49\textwidth]{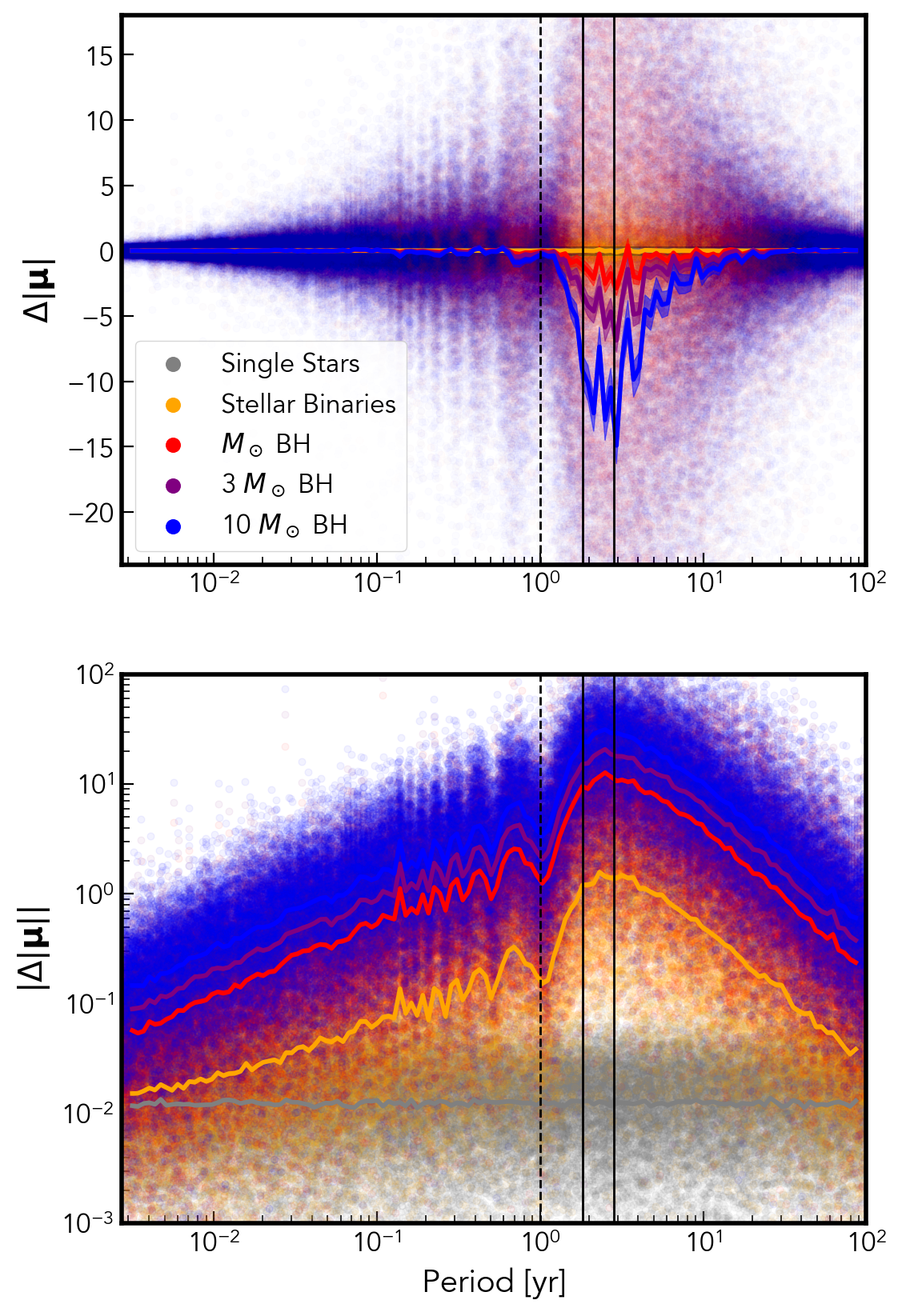}
\caption{Similar to Figure \ref{deltaMu} but now we examine the change in the (projected) speed of the system, which can take positive or negative values unlike the magnitude of the proper motion anomaly. A positive value corresponds to a source appearing to move faster in eDR3 compared to DR2.}
\label{deltaAbsMu}
\end{figure}

Changes in the proper motion of a star are another tool for inferring the presence of a companion, and has already been used in a number of surveys (e.g. \citealt{Kervella19}). If we observe a system over a fraction of its orbit then the extra motion will appear approximately linear. Then if we observe the same system again at another time that linear motion will be in a different direction (corresponding to a different phase of the orbit). 

Thus, if systems appear to have significantly different proper motions at two different times, it is strong evidence of a companion.

For a system sampled uniformly in time (a crude but useful approximation), the average proper motion caused by a binary is equivalent to the difference between the point of the orbit corresponding to the first and last observation, divided by the length of the observation window. Denoting the position of the centre of light relative to the centre of mass as \begin{equation}
\bm{\epsilon}= \frac{\varpi a \Delta (1-e^2) }{\sqrt{1-\cz_{\phi_v}^2 \sz_{\theta_v}^2} (1+e\cz_\phi)}\begin{pmatrix}\cz_\phi - \cz_{\phi_v-\phi}\cz_{\phi_v}\sz_{\theta_v}^2 \\ \sz_\phi \cz_{\theta_v} \end{pmatrix}
\end{equation}
where $\phi(t)$ is the orbital phase (see P+20 for more details), then the average proper motion is
\begin{equation}
\langle \bm{\dot{\epsilon}}\rangle=\frac{1}{t_2-t_1}\int_{t_1}^{t_2} \bm{\dot{\epsilon}}dt = \frac{\bm{\epsilon}(t_2)-\bm{\epsilon}(t_1)}{t_2-t_1}.
\end{equation}
This is not exactly the proper motion which minimises the least squares fit, but it is a close and simple approximation. For the full expression refer to Penoyre et al 2022 (in prep.).
For orbits longer than twice the observing baseline the proper motion anomaly will thus increase the longer we observe. If our observations span close to one orbit $\bm{\epsilon}(t_1)\approx\bm{\epsilon}(t_2)$ the proper motion excess will be close to 0. For shorter period objects the binary contribution becomes effectively random, but still potentially significant. If we observe over two time baselines we will find two different proper motion excesses, and can calculate the difference between them giving a proper motion anomaly.

For this discussion, we will limit ourselves to scalar change in velocity which could be expressed in full as a 2, or even 3, dimensional velocity depending on whether the system also has applicable radial velocity measurements. Even here there are two such quantities we can construct. We can examine the magnitude of the change in velocity
\begin{equation}
|\Delta \bm{\mu}| = \sqrt{(\mu_{RA^*,eDR3}-\mu_{RA^*,DR2})^2 + (\mu_{Dec,eDR3}-\mu_{Dec,DR2})^2}
\label{deltaMuEq}
\end{equation}
or the change in speed
\begin{equation}
\Delta |\bm{\mu}| = |\bm{\mu}_{eDR3}| - |\bm{\mu}_{DR2}|.
\end{equation}
The former is generally more useful, as it doesn't depend on the proper motion of the centre of mass of the system - but is limited to only positive values. The latter can have either sign. Note that we will ignore the line of sight component of the velocity in this paper, as this is an independent measurement which Gaia only provides for a subset of bright sources.

Figure \ref{deltaMu} shows the magnitude of the change of the velocity for our simulated binaries. The behaviour is broadly similar to that of UWE, increasing for larger period orbits up to the timescale of the survey, then decreasing beyond there. This is useful for further approximating the proper motion anomaly, as for these systems we expect $\langle \bm{\dot{\epsilon}}\rangle_{eDR3}\sim0$ and thus $|\Delta \bm{\mu}|\sim\langle \bm{\dot{\epsilon}}\rangle_{DR2}$. This gives a characteristic upper limit on proper motion anomaly of
\begin{equation}
|\Delta \bm{\mu}|_{max} \lessapprox \frac{\varpi \Delta a}{B_{DR2}}
\end{equation}
which becomes less relevant as we move away from periods close to $B_{eDR3}$.

There is a strong feature at periods close to 1 year, highlighting the difficulty of separating Earth's orbit from binary motion. The harmonics of multiples of the procession period of the telescope, $\sim$63 days, are also clearly visible here (they are present in all plots throughout this section but are particularly prominent in proper motion). These harmonics are only visible as we're using a realistic nominal eDR3 scanning law. They may be washed out slightly in real observations because of common small adjustments to the telescopes attitude. It is hard to differentiate if the outermost harmonic truly corresponds to the precession period or exactly half a year, perhaps both contribute.


In Figure \ref{deltaAbsMu}, we show the change in speed as a function of period. Though there is a large spread we can see that the net effect of observing the system for a longer period is to reduce the apparent speed. This is because the velocity contribution of the binary orbit is maximised if we see half an orbit, and goes to zero as we approach a full orbit. For multiple orbits this effect repeats every half orbit, but given the shorter period the semi-major axis of the orbit will be smaller (everything else unchanged) and the effect is suppressed. Thus binary stars observed over successively longer periods will appear on average to slow, tending to the proper motion of their centre of mass. This is one way in which the overlapping observational time baselines taken by Gaia differ from multiple measurements taken over distinct baselines.

In both figures \ref{deltaMu} and \ref{deltaAbsMu} we can see particular periods where proper motion anomaly is significantly suppressed. These correspond to periods for which $N_{orb}$ is close to an integer over both the $DR2$ and $eDR3$ baselines, leading to small proper motion excesses and suppressing the proper motion anomaly.

\subsection{Change in $\varpi$}
\label{changePllx}

\begin{figure}
\centering
\includegraphics[width=0.49\textwidth]{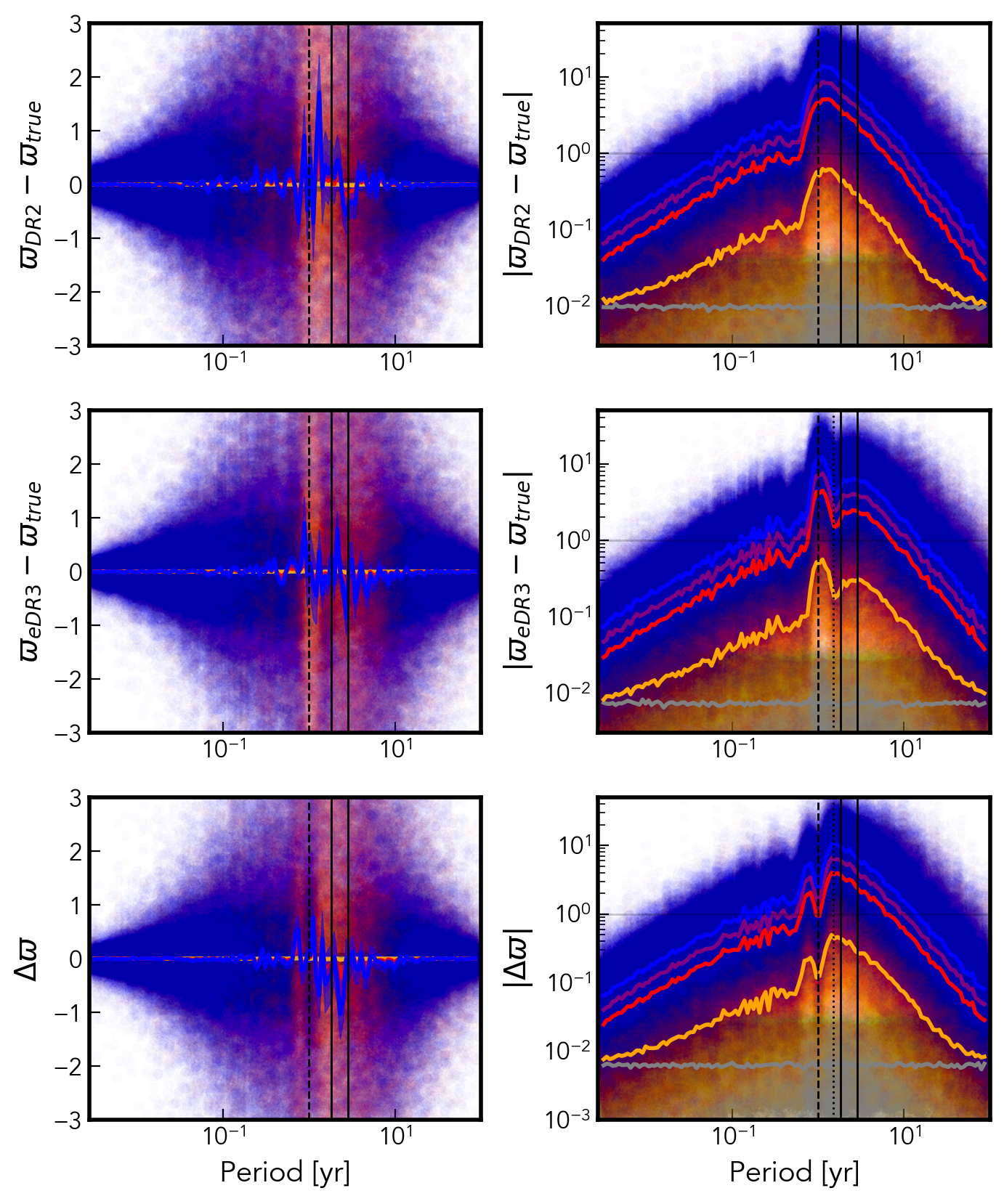}
\caption{Similar to Figure \ref{deltaUwe}, but now showing the change in parallax. In the top two rows we compare the inferred parallax from our mock observations to the true parallax (showing relative and absolute values on the left and right respectively). In the final row we show the difference in the inferred value between DR2 and eDR3. For reference a horizontal line shows a change of 1 mas in the log plots.}
\label{deltaPllx}
\end{figure}

One of the most pernicious effects of a binary is its ability to mimic a shift in parallax. Only a quite specific subset of binaries can have this effect - periods close to 1 year (or harmonics thereof), orientations aligned such that their motion mimics the parallax ellipse (itself set by the position of the source in ecliptic coordinates) and eccentricities close to zero (otherwise the motion is syncopated). Thus, the effect is likely to be rare, but with a huge sample size such as we have in the Gaia survey, rare events may occur many times. For a more detailed analytic framework see \citet{Butkevich18} and references therein.

In Figure \ref{deltaPllx}, we show the shift in inferred parallax from the true value, and the relative change in measured parallax between DR2 and eDR3. Reassuringly, we see that for most systems the shift is small, and centred on zero. Shifts of $\sim \pm 1$mas are possible for stellar binaries, and larger values for dark companions. Remembering that astrometric shift increases with parallax, some of the systems with large $\Delta \varpi$ have large parallaxes, thus the relative change is minor (rarely above $10\%$, as shown in more detail later in figure \ref{periodEverything}). That said, there will likely be a population for which their distance is significantly mis-estimated, resulting in erroneous absolute magnitudes.

It is worth noting that any parallax limited sample will contain many systems with positive $\varpi_{obs}-\varpi_{true}$ and few with a negative shift in parallax. This is because (especially within the nearest few $kpc$) the number of systems with a given parallax increases rapidly with distance. Assuming the scatter of parallax offsets is centred at 0 (as we see in Figure \ref{deltaPllx} and would expect for a linear model such as this), there will be more sources erroneously included (scattered upwards above our cutoff) than are excluded (scattered down below the cutoff). This is a version of the Lutz-Kelker bias \citep{Lutz73}, and may be the cause of some of the excess of dim sources lying below the main sequence in a Hertzsprung-Russell diagram.

Interestingly, examining $|\varpi_{eDR3}-\varpi_{true}|$ in figure \ref{deltaPllx}, we see a significant dip in the parallax error around $\frac{3}{2}$ years - which we attribute to the eDR3 baseline ($\sim 3$ years) being able to rule out this family of harmonic orbits. Thus, we may expect longer baseline observations to similarly rule out higher order harmonics, and to further reduce the contamination of binaries on parallax measurements. 

Examining the relative change of parallaxes ($\Delta \varpi = \varpi_{eDR3}-\varpi_{DR2}$), we can also see that systems very close to a 1 year period are more consistent between the two data releases than slightly longer or shorter periods, as predicted in \citet{Butkevich18}. A longer observation window would narrow this gap further, but it does show that the most pathological of systems may resist characterisation even as data continues to improve. Away from 1 year periods, the change in parallax between data releases seems like a promising way to identify erroneous solutions and even (given the very specific conditions for a binary to mimic parallax) give some insight onto the properties of the binary system.

\section{Astrometric behaviour as a function of binary properties}
\label{changeProperties}

\begin{figure*}
\centering
\includegraphics[width=0.98\textwidth]{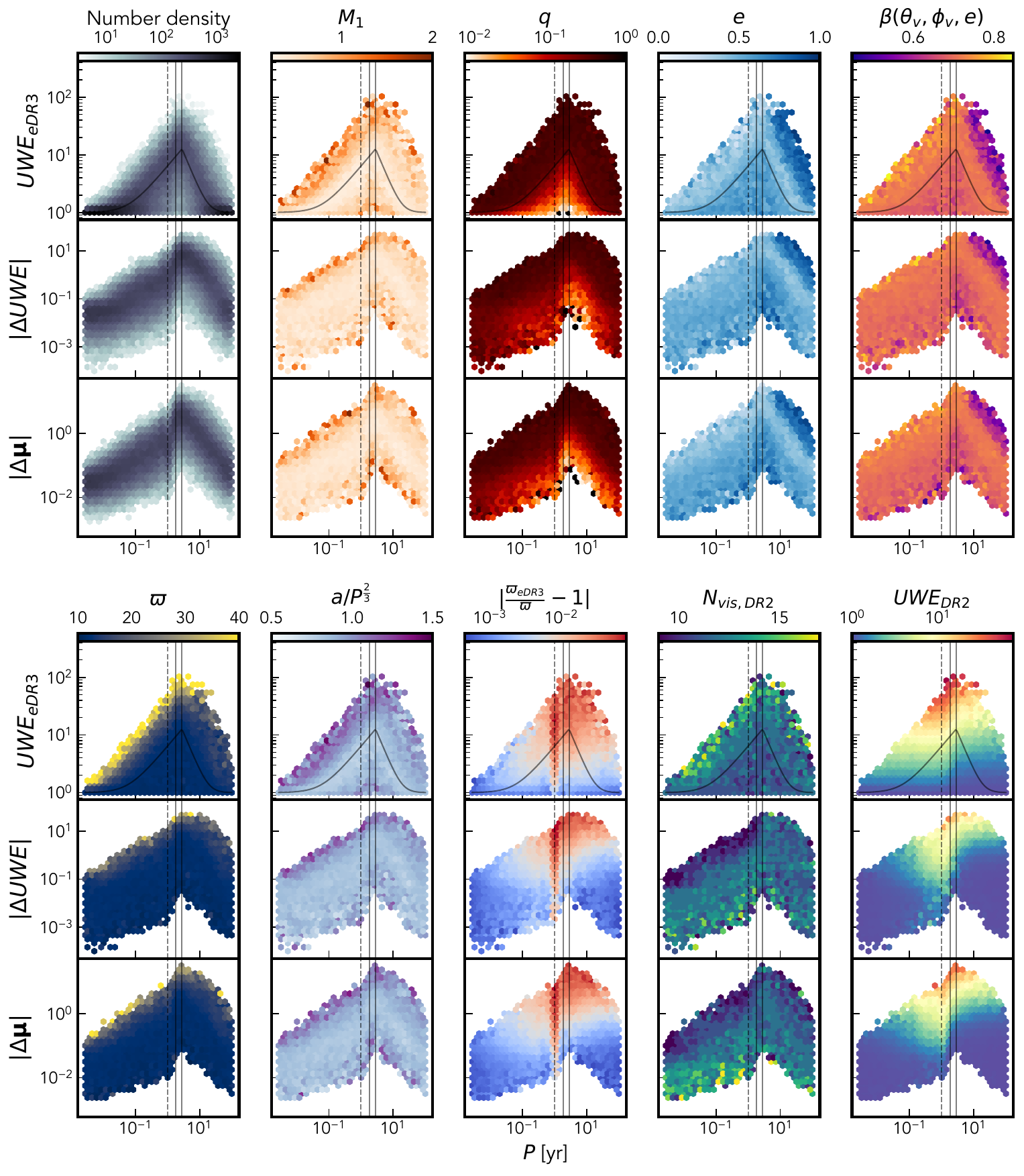}
\caption{The distribution of stellar binaries, as a function of period, of each of our astrometric quantities ($UWE_{eDR3}$, $|\Delta UWE|$ and $|\bm{\Delta \mu}|$) coloured by the various properties of the system. We show every parameter listed in table \ref{tabRandom} with the exception $l$ (which is just a rescaling of $q$), $t_{peri}$, and the position and proper motion (the former shows no dependence, and the latter mimics the behaviour on $\varpi$). The viewing angle of the binary is expressed through $\beta(\theta_v,\phi_v,e)$ (equation \ref{betaTerm}). $M_1$ is given in $M_\odot$, $a$ in AU, $P$ in years and $\varpi$ in mas. Vertical lines show periods of 1 year, 22 months (the DR2 baseline) and 34 months (the eDR3 baseline). For the $UWE$ rows we also show the predicted dependance on period from equation \ref{UWEperiod}.}
\label{periodEverything}
\end{figure*}

\begin{figure}
\centering
\includegraphics[width=0.49\textwidth]{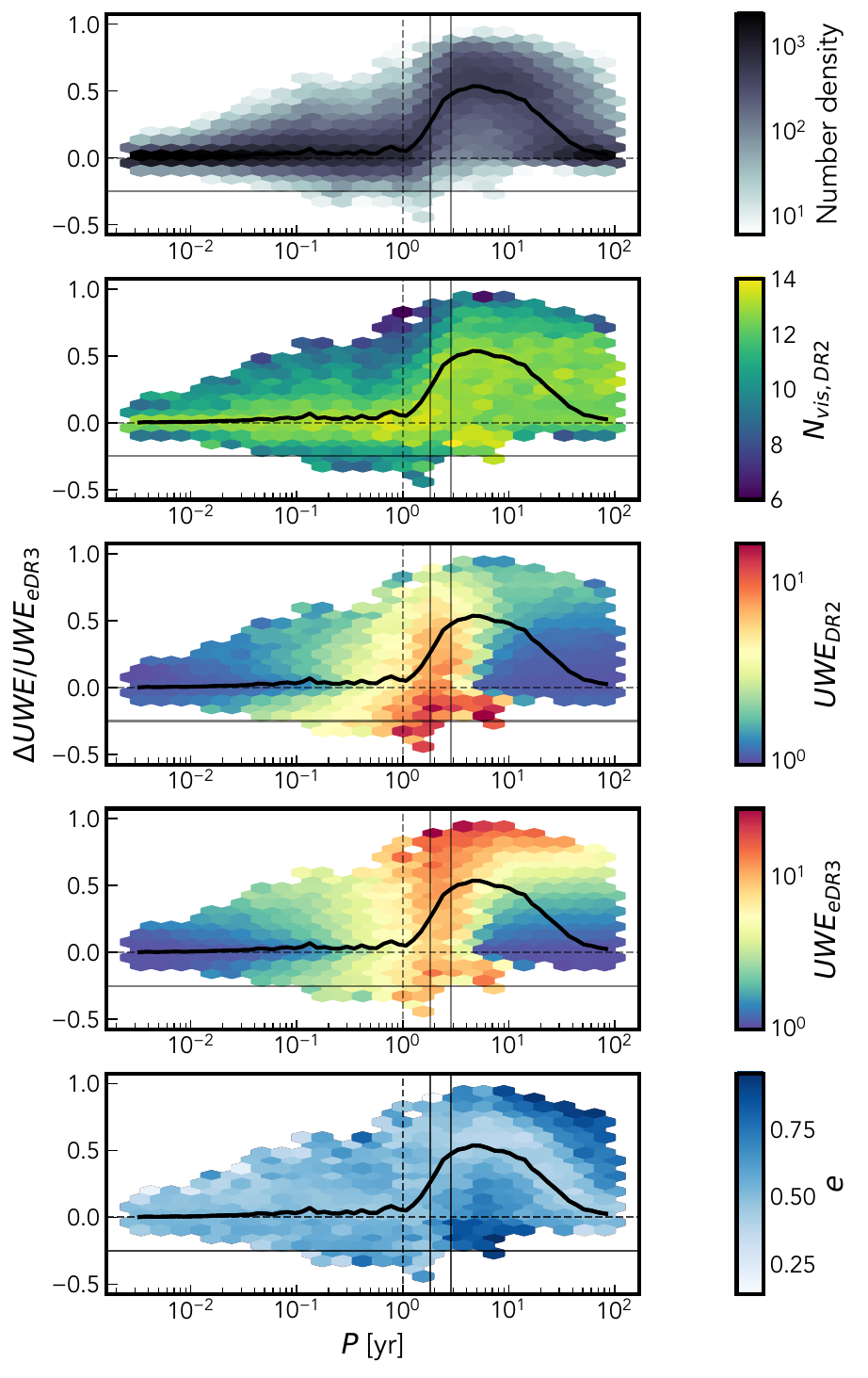}
\caption{We show the change in $UWE$, normalised by $UWE_{eDR3}$ for stellar binaries. The top panels show this behaviour as a function of period, and below we show the 1D distribution. The thick black line is the median $\frac{\Delta UWE}{UWE_{eDR3}}$ as a function of period. We also show periods of 1 year, $B_{DR2}$ and $B_{eDR3}$ as vertical lines, and $\frac{\Delta UWE}{UWE_{eDR3}}$ of 0 and -0.25 as horizontal lines.}
\label{relativeDelta}
\end{figure}

Having modelled a broad ensemble of binary properties we can examine which systems give the largest astrometric indicators. 

In figure \ref{periodEverything} we show the size of the astrometric signal for stellar binaries, as a function of period, coloured by the properties of the system.

We also show a rough sketch of the expected behaviour of $UWE$ with period which can be estimated using equations \ref{dtheta}, \ref{uwesimple} and \ref{UWEadj} and using $a\propto P^{\frac{2}{3}}$ giving
\begin{equation}
\label{UWEperiod}
UWE(P) \sim \begin{cases}
\sqrt{1+\kappa P^{\frac{4}{3}}}, & \text{for } P<B \\
\sqrt{1+\kappa B^2 P^{-\frac{2}{3}}}, & \text{for } P>B
\end{cases}
\end{equation}
where $B$ is the time baseline of observations (here $B_{eDR3}=\frac{34}{12}$)
and $\kappa$ is a constant representing the median behaviour of all other parameters, and here we use a value of 50. The period dependence is well motivated for $P<B$ but at longer periods the scaling is chosen rather arbitrarily for it's relatively good agreement with the distribution.

Going through each property in turn we'll detail how each effects the astrometric indicators and why:
\begin{itemize}
\item $M_1$ - Mass of the primary. We see here that larger mass stars give more extreme (large or small) astrometric deviations. A larger mass star leads to a larger semi-major axis for a fixed period giving the larger deviations of the upper envelope. However it also means a more extreme mass ratio, which can suppress the signal and also cause the lower envelope.
\item $q$ - Mass ratio. For most mass ratios the relative separation of the centres of mass and light scales approximately $\propto q$ (see equation \ref{deltaql}) and thus larger mass ratios give larger astrometric signals. However as we adopt $l=q^{3.5}$ mass ratios very close to one implies $q\sim l$ and $\Delta\rightarrow 0$ hence some of the smallest signals having high mass ratio.
\item $e$ - Eccentricity. We can understand the effects of eccentricity by considering it as a syncopation of time. For orbits with $P<B$ higher eccentricities give smaller signals, as these are systems where we are most likely to miss a significant fraction of the orbit from an unlucky spacing of scanning times. For $B>P$ this effect is still active, but there's also the possibility of observing a significant fraction of the orbit over a short time period if the observing window coincides roughly with periapse. Hence we see a high eccentricity fringe with large astrometric deviations at long periods.
\item $\beta(\theta_v,\phi_v,e)$ - Viewing angle (expressed via the expected dependance from equation \ref{betaTerm}). For $P<B$ we see that larger values of $\beta$ give larger signals, as we would expect. This behaviour no longer holds for $P>B$, and instead replicates the dependence on $e$ (remembering that high $e$ will give generally lower $\beta$ as seen in figure \ref{projectionTerm})
\item $\varpi$ - Parallax. The closer a system the larger the larger on-sky motion caused by a binary will appear, and hence the highest astrometric signals come from the systems with the highest parallax. The behaviour with proper motion (not shown) replicates this trend almost exactly, as for a particular true spatial velocity the on-sky motion scales linearly with parallax.
\item $a$ - Semi-major axis. The dominant dependence with semi-major axis is the trend with period ($P\propto a^\frac{3}{2}$) but by normalising with this factor we see there is some vertical structure as well. The largest astrometric signals tend to come from systems with a high $a$ compared to their period, as is especially clear in $UWE$. These correspond to more massive systems (either high primary mass, high mass ratio or both - $a\propto(1+q)M_1$ for fixed $P$). We have shown many times that the largest astrometric signals come from systems with $P\sim B$ but here we see explicitly why that is, these give the largest possible $a$ whilst still being time-resolved.
\item $|\frac{\varpi_{eDR3}}{\varpi_{true}}-1|$ - Relative parallax shift. The fitted parallax can differ significantly from the true value As we show here this especially true for systems with large astrometric deviations (at all periods) and for those with periods close to 1 year (regardless of the magnitude of astrometric deviations). The highest (median) change in parallax seen is of order 10\%, enough to add significant uncertainty to distance estimates, but not so large as to suggest parallax measurements of binary systems are inherently unusable.
\item $N_{vis,DR2}$ - Number of visibility periods over the DR2 observing baseline. The more often we observe a binary system the more significant any excess motion is over astrometric noise, and hence the highest $UWE$ is seen in systems which are best sampled by the survey. This behaviour reverses when looking at astrometric changes as these systems are dominated by those with an extreme measurement in $DR2$, which favours systems with few observations. In general the regions of the sky with few observations will be similarly undersampled (and vice-versa) over a longer baseline, and hence the behaviour with $N_{vis,eDR3}$ (not shown) is similar.
\item $UWE_{DR2}$ - Unit weight error over the DR2 baseline. In general a large astrometric deviation in $DR2$ will correspond to a similar or larger deviation when viewed over a longer baseline. In all three measures we see a marked change between $B_{DR2}$ and $B_{eDR3}$, where all binary motion observed over $DR2$ is mapped to proper motion, but the longer baseline eDR3 data is able to resolve significant curvature incompatible with straight line motion.
\end{itemize}

For most binary systems it will be the case that $l\ll q\ll1$ in which case $\delta \theta \sim q a \varpi$, which explains most of the above behaviour for $P<B$.

\subsection{Relative change in $UWE$ as a function of binary properties}
\label{relDeltaProp}

In figure \ref{relativeDelta} we also examine specifically the behaviour of $\Delta UWE$ normalised by $UWE_{eDR3}$. The median behaviour shows that the relative change is roughly, though slightly greater than, zero for long and short periods. These correspond to systems where the two $UWE$ measurements agree, either because both baselines resolve the orbit well (short periods) or not at all (long periods).

We see that the most extreme changes correspond to systems which are less well observed in the DR2 baseline. Systems with negative $\Delta UWE$ have particularly high $UWE_{DR2}$, which is somewhat reduced in $eDR3$, where the DR2 fit is dominated by a few data points with a large (but natural) scatter. We see very similar trends with eccentricity as we did in figure \ref{periodEverything}, where it can both suppress and boost $UWE_{eDR3}$ for systems with $P>B$.

\section{Predictions of observable behaviours}
\label{changeObservable}

\begin{figure*}
\centering
\includegraphics[width=0.98\textwidth]{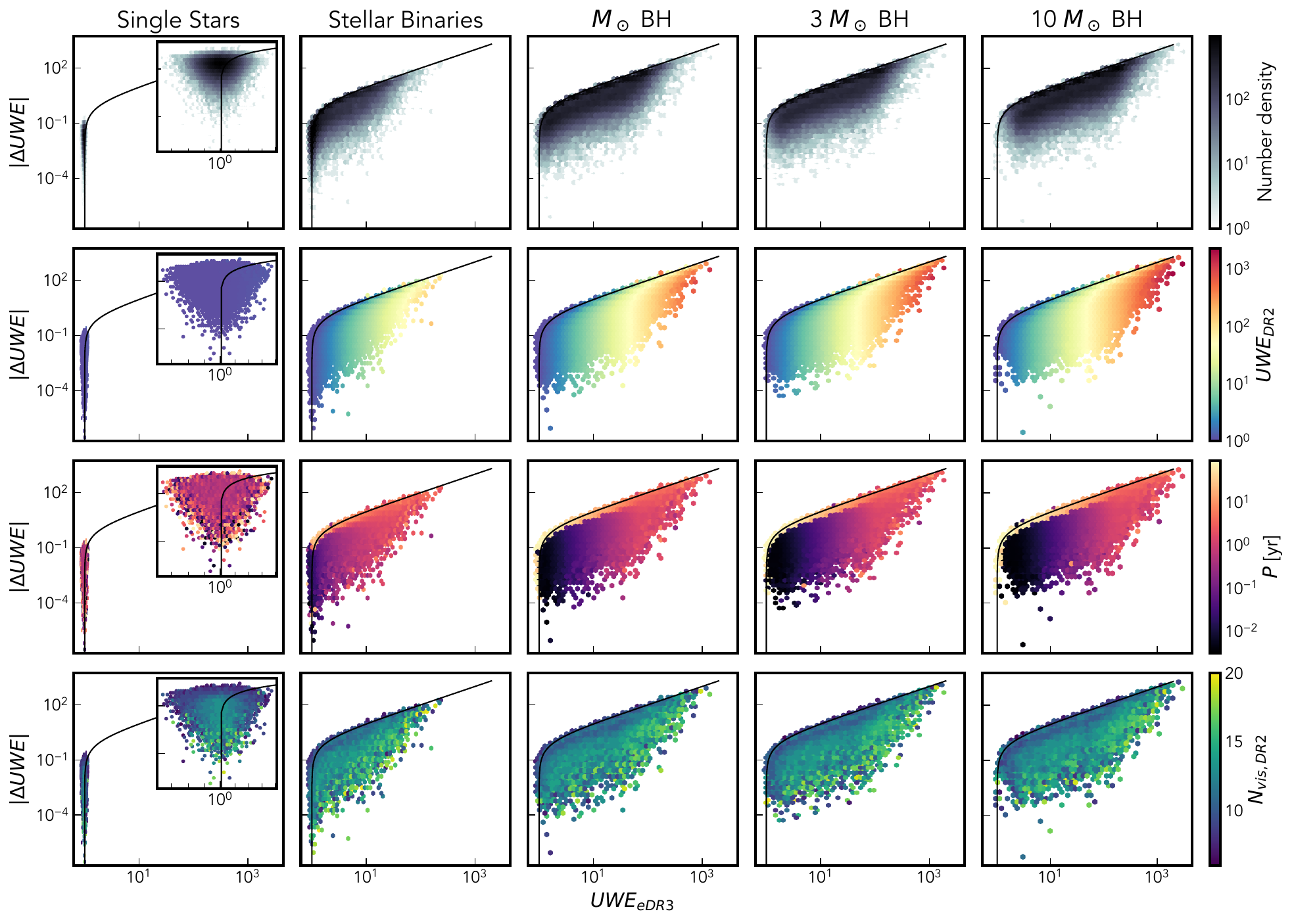}
\caption{Comparing the change in UWE to the measured UWE for each of our synthetic binary populations - as a function of the properties of the system. We show the number density (top row), measured UWE over the DR2 baseline (second row), the binary period (third row) and the number of visibility periods in the DR2 baseline (final row). Each bin shows the median properties of systems falling within it, and bins containing less than 3 systems are not shown. As single stars all have $UWE$ very close to one we show in an inset panel a zoomed view of this neighbourhood. Also shown is the line $|\Delta UWE| = UWE_{eDR3}-1$ corresponding to sources with no significant astrometric error over the $DR2$ observations, which describes the approximate upper envelope of results. Because the parameters of our systems are chosen before the companion is specified, single stars are allocated a period, though it has no impact on their astrometric track or fit (and the only behaviour seen with period is increased variance where number density is low).}
\label{ruwe_deltaruwe}
\end{figure*}

\begin{figure*}
\centering
\includegraphics[width=0.98\textwidth]{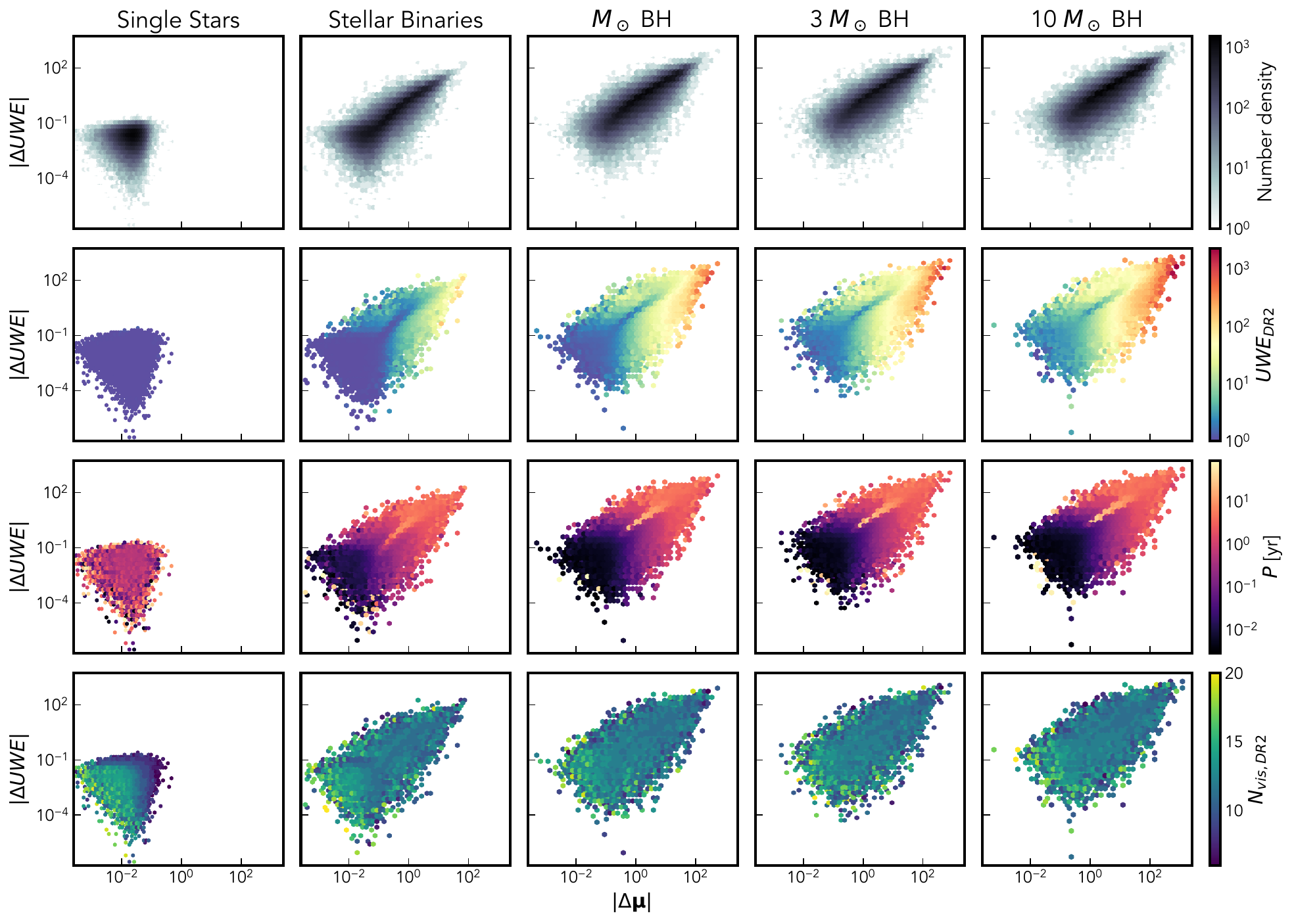}
\caption{Similar to Figure~\ref{ruwe_deltaruwe}, but now comparing the proper motion anomaly to the change in UWE.}
\label{deltapm_deltaruwe}
\end{figure*}

\begin{figure*}
\centering
\includegraphics[width=0.98\textwidth]{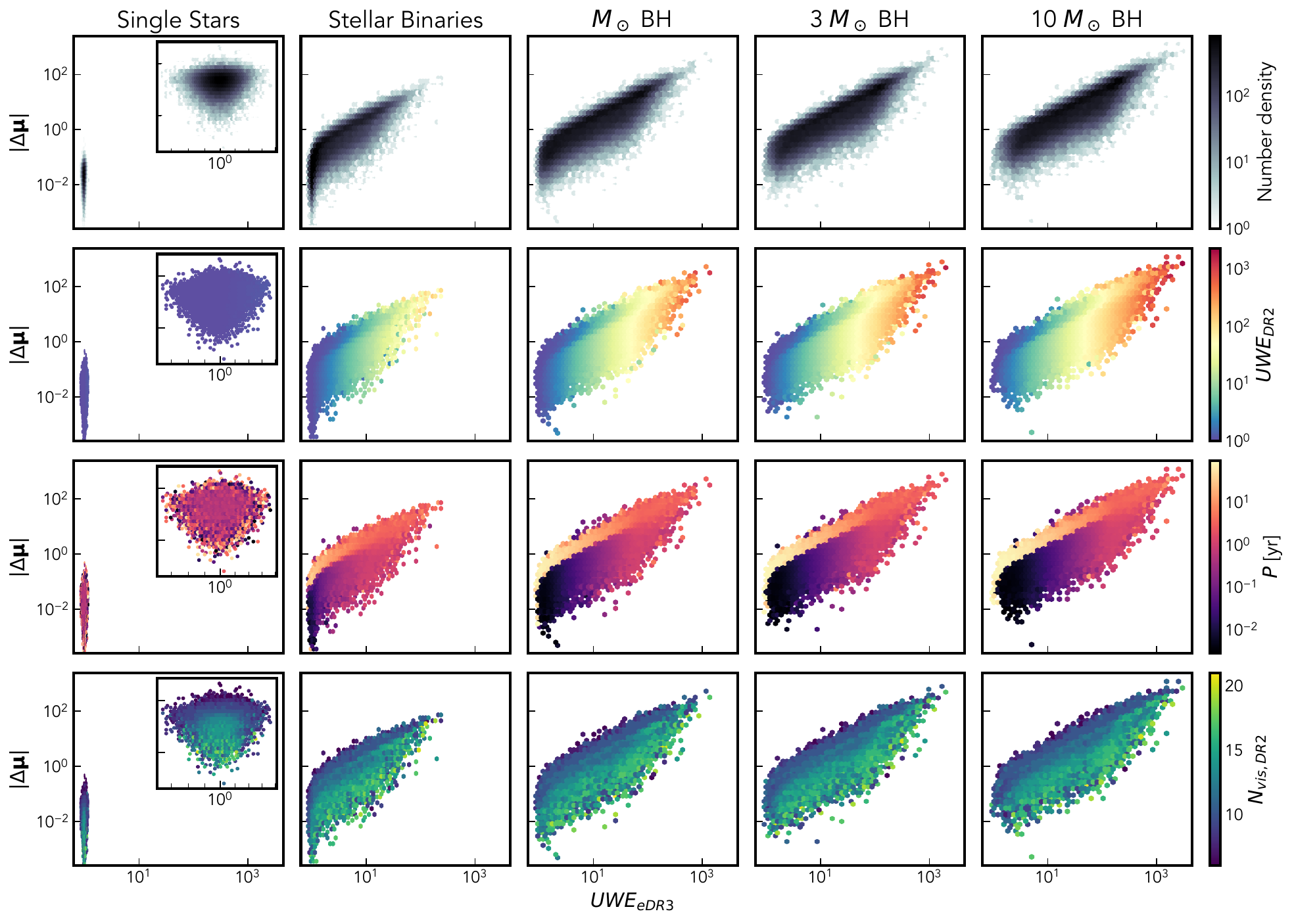}
\caption{Similar to Figure~\ref{ruwe_deltaruwe}, but now comparing $UWE_{eDR3}$ to the proper motion anomaly. Note that due to the close-to-linear relationship in figure \ref{deltapm_deltaruwe} this is almost identical to figure \ref{ruwe_deltaruwe}.}
\label{ruwe_deltapm}
\end{figure*}

Now we will move away from examining binaries and their astrometric solutions as a function of their period (which is unknown for all but a small minority of well characterised systems). Instead we ask how we should expect binaries to be distributed in terms of parameters we can infer directly from astrometric data.

We will focus on three quantities we have already introduced, $UWE$, $\Delta UWE$ (eq. \ref{deltaUWEEq}) and $|\Delta \bm{\mu}|$ (eq. \ref{deltaMuEq}). We examine both where single stars and binary systems cluster in these spaces, and the relationship between the intrinsic properties and their distribution on these planes.

Figures \ref{ruwe_deltaruwe}, \ref{deltapm_deltaruwe} and \ref{ruwe_deltapm} show the distribution and properties of our synthetic systems in the $(UWE,\Delta UWE)$, $(|\Delta \bm{\mu}|,\Delta UWE)$ and $(UWE,|\Delta \bm{\mu}|)$ planes respectively. In each, we show the overall number density, along with the dependence on the $UWE$ measured in the first observing window, the binary period and the parallax.

\subsubsection{$\Delta UWE$ vs $UWE$}
Starting with Figure \ref{ruwe_deltaruwe}, we see a very clear wing, which corresponds to astrometric fits with
\begin{equation}
|\Delta UWE| \leq UWE_{eDR3} - 1
\label{deltaUweLimit}
\end{equation}
i.e. where the $UWE_{DR2}$ measured in the first observing window was close to unity, whilst a significant $UWE_{eDR3}$ was observed over the second observing window. This corresponds to simulated sources with periods close to or above the length of the longest observing window. I.e. in DR2 only a fraction of the orbit was observed and the single body fit was $\sim$ consistent with an excess proper motion. Over the eDR3 period enough of the orbit is observed to be irreconcilable with linear motion causing a significant UWE.

We can see the general behaviour clearly when the plane is coloured by period. Very long period systems have negligible $UWE$ and $\Delta UWE$, collecting around an $UWE$ of 1, overlapping with short period (and thus small orbit) systems with negligible astrometric signals. As we move to shorter periods we trace the limit specified by equation \ref{deltaUweLimit}, with both observables peaking at $P \sim 34$ months. Below this period, $|\Delta UWE|$ is significantly reduced whilst $UWE$ decreases more gradually with period, corresponding to the decreasing semi-major axis of the orbit. This means that for a given UWE there exists two quite distinct groups of binaries, those which are and are not time-resolved. The former of which has a roughly constant $UWE$ and the latter with $UWE$ increasing the longer we observe. However, much of the clear distinction between the two groups comes from our ability here to cleanly separate binaries by mass, whereas for real data this behaviour is marginalised over the distribution of binaries of all masses and periods.

With real data unexpected noise sources will cause an extra contribution to $\Delta UWE$ which we are not capable of modelling here.



Finally examining the behaviour with $N_{vis,2}$, we see that a correlation between sources with few $DR2$ observations (and thus likely few $eDR3$ observations also) and extreme astrometric errors. This lends a lot more certainty to measurements of systems with $N_{vis,2} \gtrsim 15$.

\subsubsection{$\Delta UWE$ vs $|\Delta \bm{\mu}|$}

Moving to Figure~\ref{deltapm_deltaruwe}, we now compare the change in $UWE$ to the proper motion anomaly.

The single stars consistently have small changes in their astrometry, with the largest deviations corresponding to sources with few visibility periods in the DR2 baseline, and thus poorly constrained properties in that dataset.

For each column of binary systems, we see a significant number with low astrometric anomalies (both $UWE$ and $\Delta \bm{\mu}$), corresponding to shorter period systems for which multiple orbits are observed even during the first observing baseline. For the longest period systems, proper motion and $\Delta UWE$ are tightly correlated, again peaking around 34 months. Shorter period orbits have smaller astrometric anomalies and are less tightly correlated.


\subsubsection{$|\Delta \bm{\mu}|$ vs $UWE$}

For completeness we also show the distribution of $|\bm{\Delta\mu}|$ compared to $UWE_{eDR3}$ in figure \ref{ruwe_deltapm}. Due to the close relationship between $|\bm{\Delta\mu}|$ and $\Delta UWE$ there is little to distinguish the broad behaviours from figure \ref{ruwe_deltaruwe}.

\section{Predicted detectability}
\label{changeDetectability}

\begin{figure*}
\centering
\includegraphics[width=0.98\textwidth]{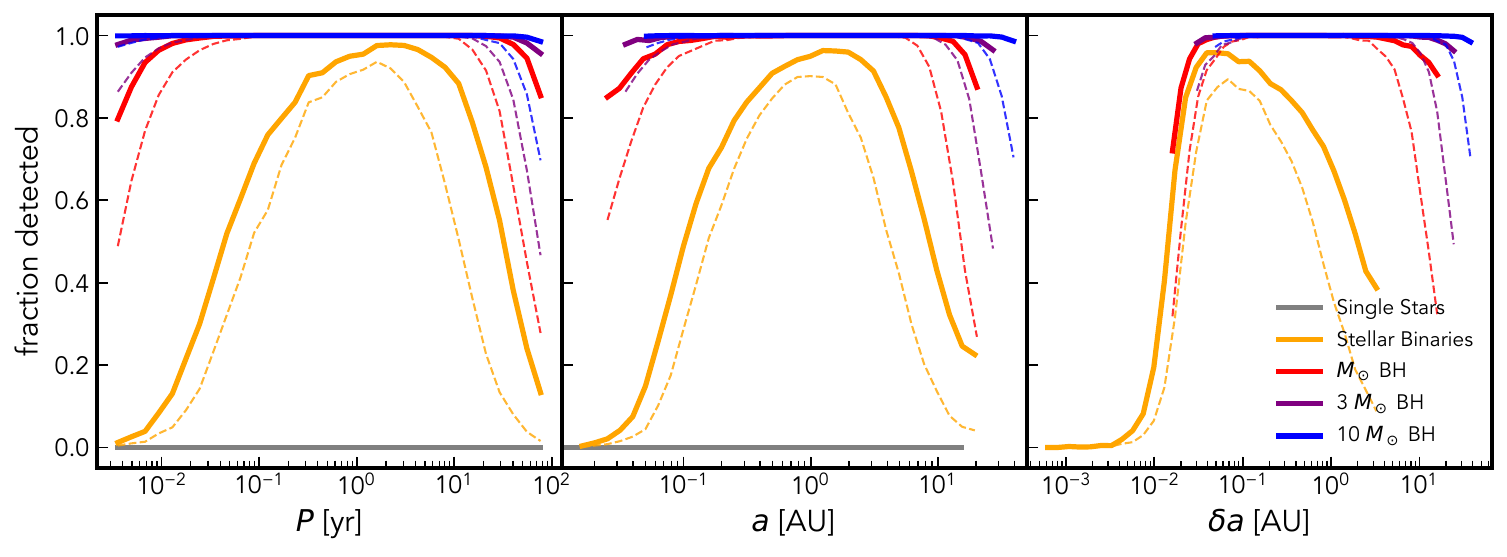}
\caption{The fraction of detected sources (those with $UWE_{eDR3}>1.25$ or $UWE_{DR2}>1.4$) for each type of source as a function of period, semi-major axis and distance between center of light and center of mass, $\delta a = \Delta \cdot a$. We use the full eDR3 observing baseline for the solid lines, and limit to just the DR2 observing window for the dashed lines. Note that single sources have $\Delta=0$ and hence cannot be plotted in $\delta a$.}
\label{detection}
\end{figure*}

\begin{figure*}
\centering
\includegraphics[width=0.98\textwidth]{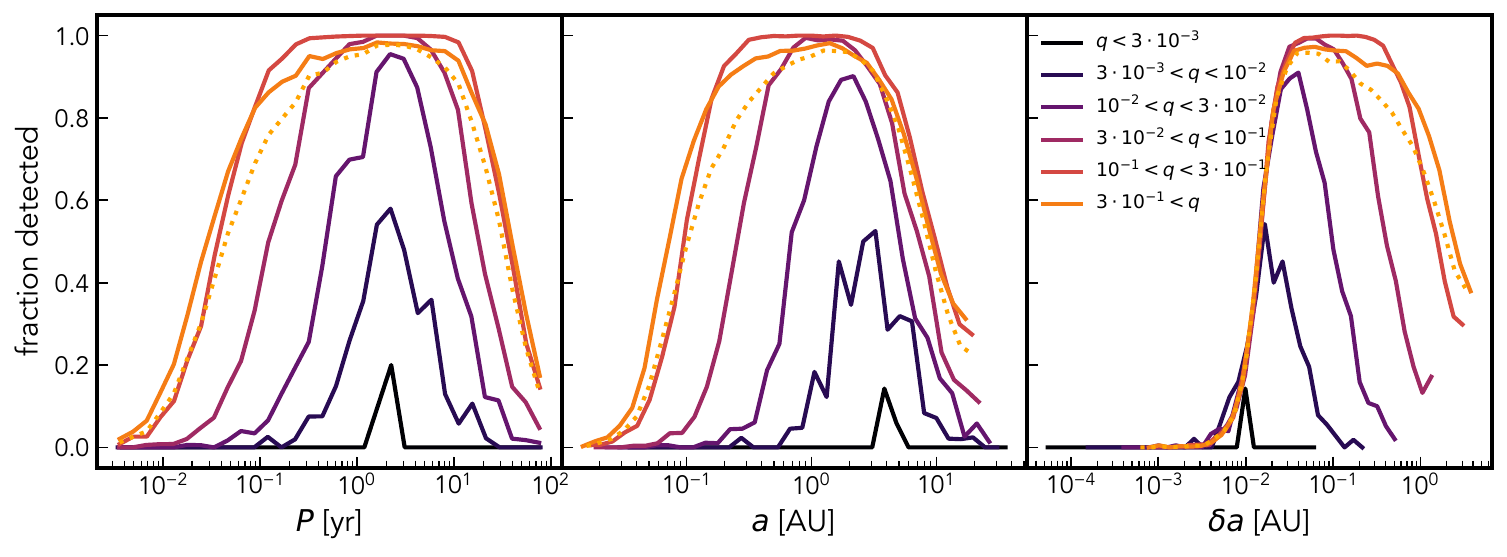}
\caption{Similar to Figure~\ref{detection}, but now only showing stellar binaries. The overall curve (the same as in the previous figure) is shown as a dotted line. The solid lines show the variation of the detection fraction (in eDR3) for sources of specific mass fractions $q$.}
\label{detection_q}
\end{figure*}

\begin{figure*}
\centering
\includegraphics[width=0.98\textwidth]{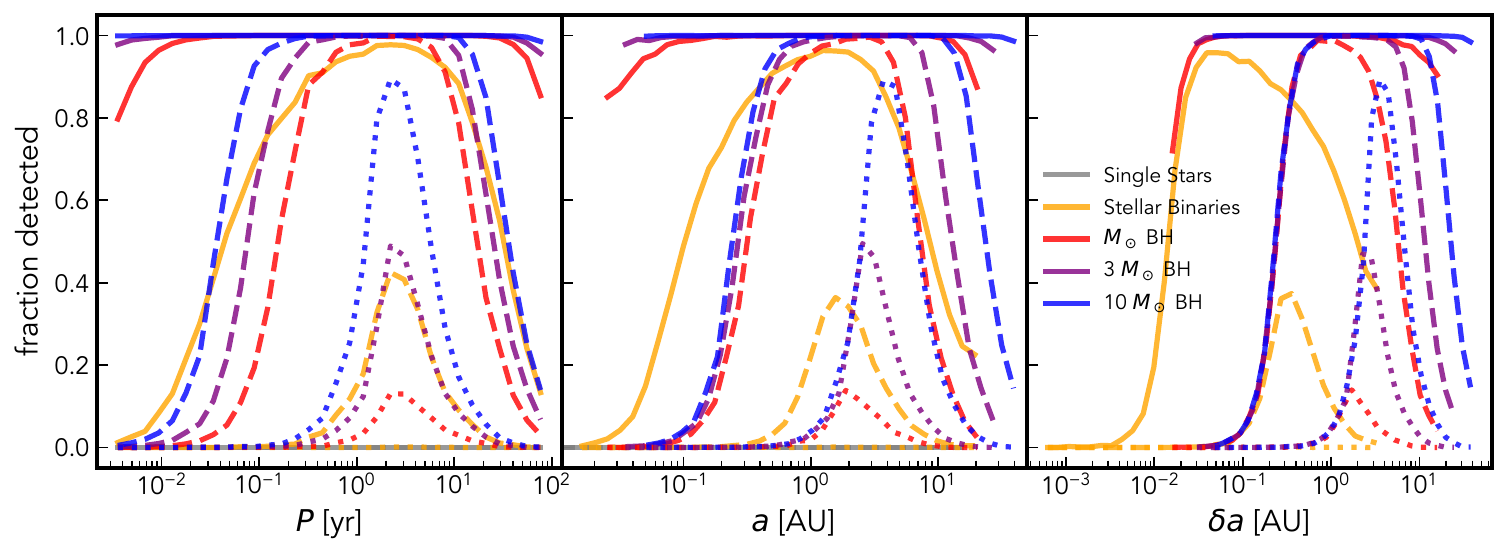}
\caption{Similar to Figure~\ref{detection} but now changing the criteria for detection. The solid lines have $UWE>1.25$ (as before), dashed lines have $UWE>12.5$ and dotted lines have $UWE>125$. This gives a rough idea of the detectability of systems at larger distances (see text for more details).}
\label{detection_distance}
\end{figure*}

The last thing we ask of our simulated systems is which we would expect to be able to detect. To do this, we will use a very simple criterion, whether a system has an $UWE_{eDR3}>1.25$ (as motivated by figure \ref{uweDist}).

Figure \ref{detection} shows the fraction of sources above this threshold for the different classes of simulated source. From this we can see clearly the range of periods and semi-major axes for which $UWE$ is an effective measure of binarity - from around a month to a decade, or 0.1 to 10 AU. Black hole hosting systems are detectable at a much wider range of periods, due to the much larger value of $\Delta$ and increase in $a$ for a fixed period. The sharp cut at low $\delta a$ shows the limit caused by the astrometric resolution (remembering that here we use a flat astrometric error of 0.1 mas for each observation, whereas in reality it varies strongly with magnitude). The improvement due to the longer time baseline in eDR3, especially for longer period wider binaries, is clear -- and similar improvements can be expected in the future data releases.

It is worth noting explicitly that the detection fraction is very high, we might expect to identify 80-90\% of binaries with periods from months to decades. The exact proportion is sensitive to the real distribution of binary properties and other observational constraints, and will in part be contaminated by other sources of astrometric noise and some small fraction of single stars (figure \ref{uweDist} shows 1 in 1,000,000 cross this boundary, but that number is likely higher for real, noisier data). However, given that binary systems are ubiquitous (between 10 and 50\% of systems are expected to host stellar companions, depending on their mass) we probably expect true binaries to dominate any selection of high $UWE$ systems and to therefore have quite well constrained periods, even before we have access to full astrometric epoch data.

Figure \ref{detection_q} shows a breakdown of just the stellar binaries in terms of the mass ratio $q$. We see that we are most sensitive to systems with $0.1<q<0.3$, with a slight reduction for larger $q$ (remembering our assumption of $l=q^{3.5}$ so as $q$ approaches unity $\Delta \rightarrow 0$). Lower mass ratio systems are detectable only close to the eDR3 observing baseline, and detections become rare for $q<0.01$. We can also see in the $\delta a$ panel that detectable systems pile up at the resolution limit, especially at lower mass ratios. Thus we might expect to be able to detect lower mass ratio systems for sources with lower astrometric error (e.g. an apparent magnitude of $\sim$ 13 where Gaia's astrometric error per CCD is $\sim$0.2 mas).

We have modelled every system we simulated with a black hole companion, mostly as an example of what we expect to see from an extreme mass and light ratio system. However, taken literally, this would suggest that for every nearby single star or stellar binary there is also a black hole of significant mass, which is certainly not true. We might expect a few black holes of $1 M_\odot$ within 100's of pc, and BHs of $10 M_\odot$ to be at distances of $1-10$ kpc \citep{Chawla21}. 

The average angular seperation of a binary, $\delta \theta$, is proportional to the parallax of a system, and $UWE \rightarrow \delta \theta$ for $UWE \gg 1$. Thus, for a given binary, our ability to resolve the motions of these black hole hosting binaries depends directly on the expected distances.

We could resimulate our BH binary systems with a more realistic distance distribution (c.f. the parameters used in table \ref{tabRandom}), but there is a simple way to probe the same question without new data: increasing the critical $UWE$. If we for example ask which systems have $UWE_{eDR3}>125$ (compared to 1.25) we are effectively asking which systems would still be visible 100 times further away, much more in line with the expected distances to real BHs. Thus in figure \ref{detection_distance} we show the fraction of sources which meet $UWE$ cuts 10 and 100 times larger.

We see the period dependence tighten on our optimal period of $B_{eDR3}$ with stellar binaries dropping to zero at large distances. Some BHs are still visible at 10 kpc scales (corresponding to $UWE>125$) and their visiblity is now very dependant on their mass (which in turn sets $a$). We see the range of detectable semi-major axis shift to larger orbits, the only systems where $\delta \theta$ still dominates of $\sigma_{ast}$.

The validity of this calculation also relies on our flat $0.1$ mas astrometric error still remaining relevant. As astrometric error depends on apparent magnitude (increasing rapidly beyond $\sim15$) this will further select for the brightest stars, and makes our rough estimate into a likely upper limit.

The heartening result from this simple analysis is that the detection fraction of astrometric BHs is significant even at large distances. This is in line with more careful predictions of the detectability of massive BHs via astrometry \citep{Chawla21}. The remaining challenges will be in successfully differentiating BH host systems from other sources of astrometric noise.

\section{Conclusions}

In this paper, we explore how deviations from single object astrometry allow us to infer the presence of an unresolved binary companion. Specifically, we look at our ability to identify individual binary systems, based on datasets spanning the Gaia DR2 and eDR3 epochs, by simulating binary systems and fitting them with a close emulation of \textit{Gaia's} own astrometric pipeline. We limit our analysis to nearby systems, within 100 pc, and in paper II of this series we will present the corresponding measured astrometric deviations from the Gaia Catalog of Nearby Stars \citep{Smart20}.

We start, in Section \ref{syntheticbinaries}, by drawing binary parameters from some simple distribution and generating astrometric tracks. We sample these based on the nominal Gaia scanning law and fit with a direct emulation of the satellite's own astrometric data processing procedure. We do this for both the Gaia DR2 and eDR3 observing baselines (the latter of which encompasses and extends the former) giving two sets of astrometric measurements for each system, which we can then compare. 

In Section \ref{changePeriod} we show that in general the Unit Weight Error (a measure of goodness-of-fit which should take values close to 1 for a single source) is consistent between measurements for systems with periods shorter than the DR2 observing window. Systems with longer periods, above the observing baseline for DR2, in general show an increase in UWE.

This provides a valuable method for separating true binary contributions (which will either increase UWE or leave it unchanged) from other sources of astrometric contamination (for which the change in UWE can take either sign). We suggest that sources with $\frac{\Delta UWE}{UWE_{eDR3}}<-0.25$ should be considered inconsistent with both single stars and most binaries, and such a cut should highlight significant numbers of spurious astrometric solutions not caused by binarity.

Our single stars have a tighter distribution of $UWE$ in eDR3 compared to DR2, mostly stemming from a higher number of observations. Thus we suggest that the critical $UWE$ above which we should consider a source inconsistent with a 5-parameter single body motion should be reduced for eDR3 (and again in future longer baseline data releases). By comparing the distribution of DR2 and eDR3 single star UWEs we suggest an analogous criteria is $UWE_{eDR3}>1.25$. Around 60\% of our simulated stellar binaries are above this criteria, compared to $\sim$45\% with $UWE_{DR2}>1.4$.

We also examine the change in proper motion and parallax in terms of binary period, both of which show a similar change in behaviour at the length of the eDR3 observing window.

Proper motion anomalies (PMA) are largest in systems with periods that match (or are just below) the eDR3 time baseline. At longer periods the PMA reduces, but less sharply than $UWE$, confirming that PMA is sensitive to slightly longer period systems, whilst less sensitive for periods $<1$ year. We also show that on average binary sources appear to slow over time, as one or more whole orbits are observed.

As described in P+20, we see a significant parallax anomaly (compared to the true value) for binary systems. This is especially large in systems with periods close to a year, but also for those with the largest astrometric deviations overall. The magnitude of the shift is generally 10\% or less, making the measurement of the distance to binary systems uncertain but not overly unreliable.

In Section \ref{changeProperties} we show how the other properties of a binary effect the astrometric signals. We show that increased mass ratio or parallax leads to a larger signal, whilst other parameters like the eccentricity, viewing angle and number of observations cause more complex but significant variations. Some parameters, such as the on-sky position and proper-motion (once we've accounted for parallax) have negligible impact on the fits.

Much of the above analysis focuses on the behaviour as a function of period - a crucial predictor of behaviour, but not a quantity we know or can easily estimate for most binary systems. Thus in section \ref{changeObservable} we compare the distribution of single stars and binaries as a function of astrometric measurements that \textit{Gaia} can measure. 

We show that the proper motion anomaly and change in UWE are highly correlated for binary systems, and thus the behaviour of either as a function of $UWE$ 
closely resembles the other. 

When comparing astrometric changes to $UWE$ we see that systems with marked change between DR2 and eDR3 correspond mostly to longer binary periods. There is also a strong dependence on the number of viewing periods (well spaced astrometric measurements) with less sampled systems having in general larger variation in astrometric measurements. Astrometric measurements, especially of binaries, are significantly more reliable for higher $N_{vis}$. We suggest $N_{vis,2}>15$ as an indicator of a well sampled system, but do not suggest using this as a general criteria as, due to \textit{Gaia's} scanning law, this effectively selects particular regions of the sky.

Finally in section \ref{changeDetectability} we ask what fraction of binary systems would be detected by a simple $UWE_{eDR3}>1.25$ cut, as a function of period. We show that for a wide range of periods (months to decades) more than 80\% of all simulated stellar binaries would be selected. Using the equivalent metric in DR2 we would find 5-10\% less systems and be limited to a narrower period range, being especially insensitive at longer periods.

We also split our stellar binaries by mass ratio, and show that our sensitivity drops sharply for $q$ of $10^{-3}$ or below, and those systems which are detected have periods much more tightly peaked at the eDR3 baseline of 34 months. This directly informs which low mass companions (primarily brown dwarfs and exoplanets) we will be able to detect with current observations. Close analogs of Jupiter are likely to give observable signatures, and the detection fraction around lower mass stars is likely to be significantly higher (this result agrees well with, for example \citealt{Holl21}). 

Given our simple assumption about the luminosity ratio of stars ($l=q^{3.5}$) systems with q very close to one actually become slightly harder to detect, as the photocenter and center of mass both sit almost exactly halfway between the two stars. However real systems are likely to have more natural variation than our simple model allows, and thus this may not be as reliably true for observed binaries of near identical stars.

Lastly we show a simple approximation to the fraction of systems we'd detect at greater distances. We do this by raising the critical $UWE$ by some factor, which would correspond to detections with the same efficiency at a  distance increased by the same factor. Our simulated black hole hosts are at an unphysically high local density in the original sample (and are intended more as examples of the effect of extreme mass ratios than likely observations) but we see that at more reasonable distances the $UWE$ caused by a dark companion should still be observable for favourable periods.

In conclusion we have shown that the extended time baseline of \textit{Gaia} eDR3 has significantly improved the reliability and range of detectable astrometric binaries. As well as more robust data we also gain the ability to compare astrometric measurements from multiple data releases to further discriminate between true multiple body systems and other sources of noise. This improvement is likely to apply just as much to future data releases, with yet longer baselines. 

We have shown that astrometric deviations such as the $UWE$, change thereof and proper motion anomaly are very effective separating out binary systems with periods ranging from months to decades, from single stars. We are able to confirm the formula presented in P+20 and show that these results are relatively insensitive to the added complexity of a realistic 1D scanning law.

The systems which show large signals here are also those which, with full epoch astrometric data (as anticipated from future data releases) are most amenable to fitting full binary models. Thus we can see the breadth and properties of systems for which Gaia will give us full orbital solutions.

We will continue the spirit of this investigation in paper II, where we turn to the real sample of \textit{Gaia} observations within 100 pc and will compare our predicted behaviour with the empirical behaviour, and produce a catalog of candidate astrometric binaries close to our Sun.

\section*{Data Availability}

The data underlying this article is freely available at \url{https://zenodo.org/record/6412811}. All data was generated using the open-source code \texttt{astromet.py} which is available at \url{https://github.com/zpenoyre/astromet.py}.

\section*{Acknowledgements}

We would like to thank Berry Holl, the referee, whose detailed and careful comments and questions have led to innumerable improvements to the work. We thank the Cambridge Streams group, including Andrew Everall, Sergey Koposov, Semyeong Oh, Jason Sanders, Leigh Smith, Shion Andrews and Eugene Vasiliev for their comments, questions and contributions throughout the preparation of this work. We would also like to thank the referee for their detailed and significant suggestions and improvements. ZP would also like to thank Emily Sandford for her help preparing the draft. This work has made use of data from the European Space Agency (ESA) mission
{\it Gaia} (\url{https://www.cosmos.esa.int/gaia}), processed by the {\it Gaia}
Data Processing and Analysis Consortium (DPAC,
\url{https://www.cosmos.esa.int/web/gaia/dpac/consortium}). Funding for the DPAC
has been provided by national institutions, in particular the institutions
participating in the {\it Gaia} Multilateral Agreement.

\bibliographystyle{mnras}
\bibliography{bib}
\bsp

\appendix

\section{Dependence of UWE distributions on stellar binaries properties}
\label{sec:uweDist_properties}

\begin{figure}
\centering
\includegraphics[width=0.49\textwidth]{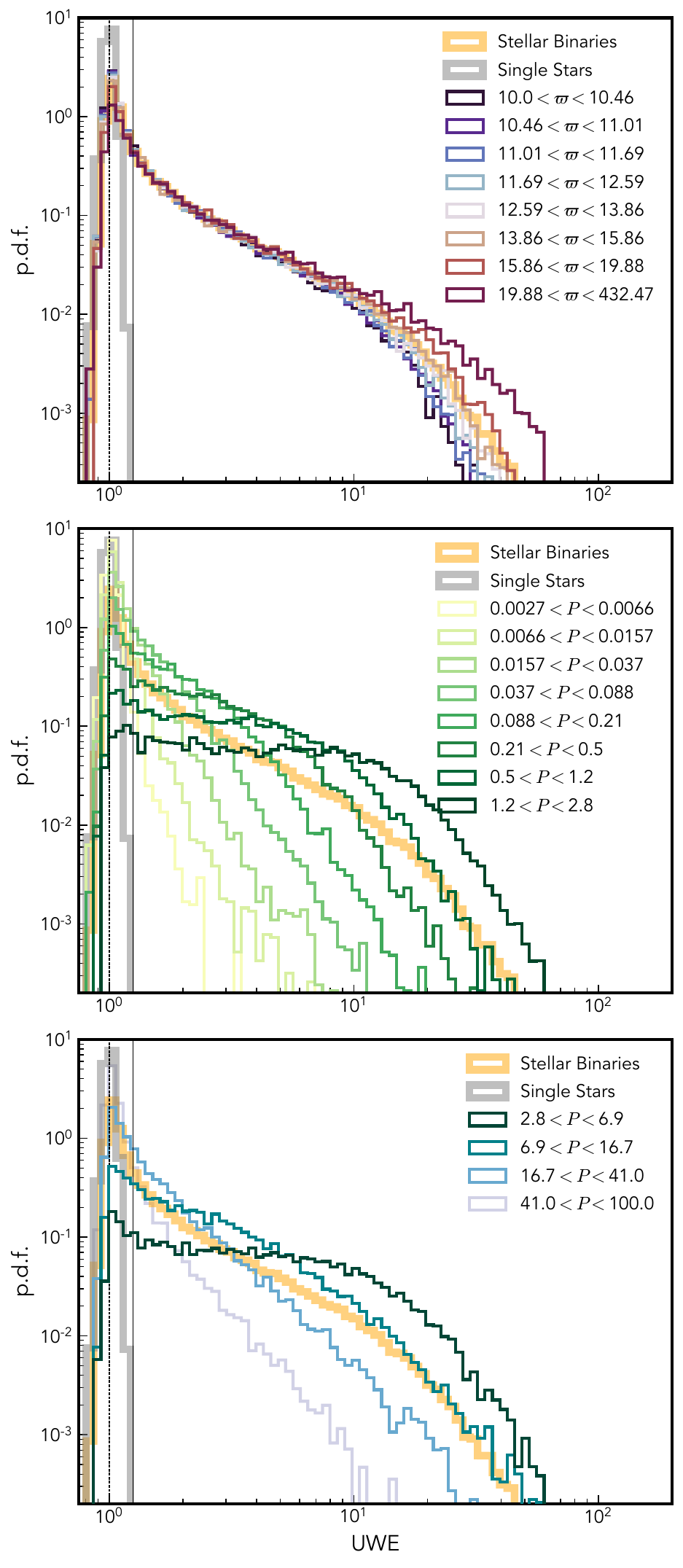}
\caption{Similar to figure \ref{uweDist} but now separating the stellar binaries by their properties (into groups with equal numbers of members. In the top panel we show the behaviour with parallax. In the middle and bottom panels we show the separation based on binary period, which we split into those with periods less than (middle) and greater than (bottom) the \textit{Gaia} eDR3 time baseline.}
\label{uweDist_properties}
\end{figure}

In figure \ref{uweDist_properties} we split the stellar binary population into sub-populations of particular parallax and period.

The reason we focus on this space in particular is that in many cases it is the first, and often the most reliably informative way to examine the astrometric deviations of a sample. As we will argue in paper II the criteria for selecting binary sources should be informed directly from the distribution of $\log(UWE)$. The peak should be easily visible for most populations, at or close to an $UWE$ of 1. The width of the $UWE=1$ peak will effect the particular $UWE$ criteria used, with a wider peak necessitating a larger critical $UWE$ and thus a lower number of binary sources/ higher degree of binary contamination. Depending on the sample the high $UWE$ tail may take different forms or even be negligible, in which case astrometric error alone is insufficient for selecting binaries.

The criteria that we suggest in this paper, of $UWE_{eDR3}>1.25$, is likely a best case scenario - our simulated data is missing many possible noise sources (as detailed in section \ref{changeInUwe}) - and thus real datasets may require a higher $UWE$ cut. Any extra spread in $UWE$ in the single source peak will likely also extend to astrometric binaries, but sources with large $UWE$ will still be distinguishable.

Understanding how the properties of a population of binaries effects its distribution of $UWE$ can inform us of which binaries we are likely to miss if the critical $UWE$ is increased, and is the first step to being able to infer the binary fraction and properties directly from the observed $UWE$s. 
Starting with the first panel of figure \ref{uweDist_properties} we see that higher parallax (and thus closer) sources have a larger $UWE$ in general, as expected from equation \ref{uwesimple}. As these sources are uniformly distributed in 3D space the vast majority are close to the distance cutoff (e.g. half of our sources with 100 pc are further than 75 pc away) and thus the effect is quite slight. This obviously isn't the case if we're interested in particular systems, where the closest may yield observational effects of interest on too small an angular scale to be seen further away. It is interesting to note however that the number of intermediate $UWE$ (between $\sim$2 and 10) sources is almost constant between distance bins, it is just the high $UWE$ end of the distribution that's particularly affected.

As we have shown throughout this paper the largest $UWE$ population is that with binary periods close to the time baseline of the survey (here 34 months for eDR3) and at much lower or higher periods the distribution becomes much more steep and tends towards the single star behaviour. Even at a fixed period the distribution of $UWE$ is still dependant on other choices of parameters, most notably the stellar masses which directly affect the semi-major axis, and to a smaller extent the eccentricity. With large enough samples of binaries of known period we may be able to constrain the distribution of these parameters from the $UWE$ distribution alone.

\label{lastpage}

\end{document}